\documentclass[12pt]{iopart}

\usepackage{iopams}  
\usepackage[pdftex]{graphicx}					
\usepackage[latin1]{inputenc} 
\usepackage[T1]{fontenc} 
\usepackage[english]{babel}

\renewcommand{\b}[1]{\mathbf{ #1}}									

\renewcommand{\d }{\mathrm{d}}										
\newcommand{\desude}[1]{\frac{\d }{\d #1}}						

\newcommand{\ket}[1]{\mid\!#1\rangle}						
\newcommand{\cop}[1]{#1^{\dagger}\!}								
\newcommand{\aop}[1]{#1}											
\newcommand{\bok}[3]{\langle #1 \! \mid \! #2 \! \mid  \! #3  \rangle}	
\newcommand{\m}[1]{\langle #1 \rangle}								
\newcommand{\rom}[1]{\uppercase\expandafter{\romannumeral #1\relax}} 
\newcommand{\id}{\hat{1}}										
\newcommand{\operatorname}[1]{\mathrm{#1}}
\newcommand{\odd}{\mathrm{o}}	
\newcommand{\even}{\mathrm{e}}											
\newcommand{\h}[1]{\hat{#1}}			

\begin{document}

\title[Collective dynamics of multimode bosonic systems]{Collective dynamics of multimode bosonic systems induced by weak quantum measurement}

\author{Gabriel Mazzucchi}
\address{Department of Physics, Clarendon Laboratory, University of Oxford, Parks Road, Oxford OX1 3PU, United Kingdom}
\ead{gabriel.mazzucchi@physics.ox.ac.uk}
\author{Wojciech Kozlowski}
\address{Department of Physics, Clarendon Laboratory, University of Oxford, Parks Road, Oxford OX1 3PU, United Kingdom}
\author{Santiago F. Caballero-Benitez}
\address{Department of Physics, Clarendon Laboratory, University of Oxford, Parks Road, Oxford OX1 3PU, United Kingdom}
\author{Igor B. Mekhov}
\address{Department of Physics, Clarendon Laboratory, University of Oxford, Parks Road, Oxford OX1 3PU, United Kingdom}
\vspace{10pt}
\begin{indented}
\item[]\today
\end{indented}

\begin{abstract}
In contrast to the fully projective limit of strong quantum measurement, where the evolution is locked to a small subspace (quantum Zeno dynamics), or even frozen completely (quantum Zeno effect), the weak non-projective measurement can effectively compete with standard unitary dynamics leading to nontrivial effects. Here we consider global weak measurement addressing collective variables, thus preserving quantum superpositions due to the lack of which path information. While for certainty we focus on ultracold atoms, the idea can be generalized to other multimode quantum systems, including various quantum emitters, optomechanical arrays, and purely photonic systems with multiple-path interferometers (photonic circuits). We show that light scattering from ultracold bosons in optical lattices can be used for defining macroscopically occupied spatial modes that exhibit long-range coherent dynamics. Even if the measurement strength remains constant, the quantum measurement backaction acts on the atomic ensemble quasi-periodically and induces collective oscillatory dynamics of all the atoms. We introduce an effective model for the evolution of the spatial modes and present an analytic solution showing that the quantum jumps drive the system away from its stable point. We confirm our finding describing the atomic observables in terms of stochastic differential equations.
\end{abstract}
\maketitle

\section{Introduction}
Quantum measurement is one of the most intriguing aspects of quantum mechanics. Many of its unusual manifestations have been already demonstrated, which includes quantum jumps and quantum Zeno effect. In particular, intriguing quantum states of photons such as Fock and Schr\"odinger cat states were prepared using measurement backaction \cite{HarocheBook,Vlastakis2013}. In a typical scenario, continuous measurement leads to projection to a well-defined state. Here, we focus on an essentially many-body (or multimode) version, where there are many particles (modes) indistinguishable by the measurement. Thus the measurement is collective and preserves quantum superposition for indistinguishable emitters. In addition, the measurement is weak, thus it can compete with the unitary dynamics \cite{Mazzucchi2016}. We discuss the result of such a competition. Although for certainty we consider ultracold atoms trapped in optical lattices, the idea can be applied to other arrays of quantum emitters, e.g. superconducting qubits as used in circuit cavity QED \cite{Hartmann2006,Palacios2010,Paraoanu2011,Pirkkalainen2013,White2015,Bretheau2015}, matter waves scattering \cite{Mayer2015}, Rydberg \cite{Zeiher2016,Schauss2015} and other polaritonic and spin excitations \cite{Budroni2015}, optomechanical arrays \cite{Paternostro2011,Aspelmeyer2014}, multimode cavities \cite{Gopalakrishnan2009,Kollar2015}, 
and even purely photonic systems with multiple path interference (where, similarly to optical lattices, the quantum walks and boson sampling were discussed \cite{Spring2013,Nitsche2016,Brecht2015,Elster2015}).

Ultracold gases trapped in optical lattice are an extremely flexible tool for studying the behaviour of matter in the degenerate regime \cite{Lewenstein}. The possibility of realizing  a vast range of Hamiltonians with highly tunable parameters make these systems suitable for studying phenomena from different disciplines, from condensed matter to particle physics.  Recent experimental  breakthrough~\cite{Hemmerich2015,Esslinger2015} succeeded in coupling these systems to optical cavities, opening the possibility of merging the field of ultracold gases to quantum optics and realizing the ultimate regimes where the quantum nature of both matter and light is equally important \cite{Mekhov2012,ritsch2013}. These new experiments enable the study of novel effects where the atomic interactions are mediated by the light field \cite{Moore1999, Chen2009, Caballero2015, Caballero2015a,Caballero2016bond}, enriching the phenomenology of these systems and leading to new quantum phases. Moreover, the entanglement  between the light and matter degrees of freedom is at the core of several quantum nondemolition proposals~\cite{Mekhov2009PRA, Szigeti2009,Roscilde2009, Rogers2014, Eckert2007, HaukePRA2013, Rybarczyk2015, Atoms,Kozlowski} where atomic properties are inferred from the observation of the scattered light. 

We focus on the measurement backaction on the atomic state due to the detection of the photons escaping the cavity. This effect competes with the usual atomic dynamics and can be exploited for engineering interesting quantum states (with \cite{Szigeti2010,Hush2013,Pedersen2014,Ivanov2014,Ivanov2016,Mazzucchi2016fb} or without \cite{Mazzucchi2016, Mazzucchi2015} the need of external control), tuning quantum Zeno dynamics~\cite{Kozlowski2015NH} and tayloring the phase diagram of the system \cite{Mazzucchi2016,Ashida2016}. We consider a spatially structured measurement scheme which partitions the system in multiple spatial modes with non-trivial overlap \cite{Elliott2015} and we focus on the evolution of the atomic wavefunction in a single experimental run. We formulate an effective description of the dynamics of the system in terms of collective variables characterizing each spatial mode. Depending on the spatial structure of the measurement operator, the detection process can induce induce dynamical multi-mode macroscopic superposition states which preserve long-range coherence. We present an analytically solvable model valid in the case of two spatial modes which shows how the quantum jumps drive the atomic system away from its stationary point,  capturing the emergence of large-scale collective oscillations with increasing amplitude.  We confirm our findings by studying the stochastic differential equations that govern the evolution of the atomic observables.

\begin{figure}[t]
\centering
\includegraphics[width=0.5\textwidth]{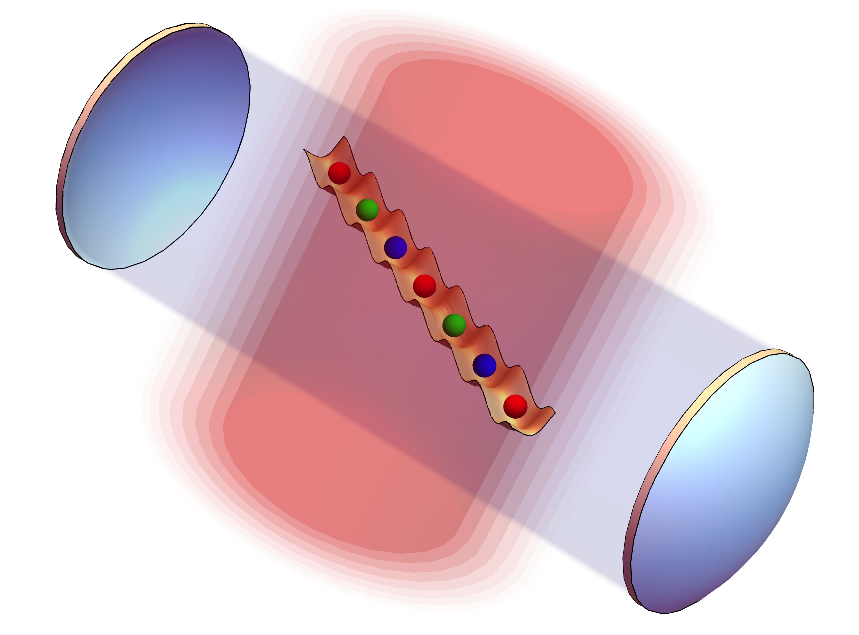}
\caption{\label{setup} Setup. An optical lattice is probed with a  coherent light beam (red) and the scattered light is enhanced by an optical cavity (blue). Depending on the angle between the two beams, the measurement defines macroscopically occupied spatial modes which maintain long-range coherence (represented by atoms of the same color).}\label{fig:setup}
\end{figure}

\section{Theoretical Model}

We consider a system of $N$ ultracold bosons loaded in an optical lattice with $M$ lattice sites that scatter light into a cavity with cavity decay $\kappa$ (Figure~\ref{fig:setup}). The cavity enhances the light scattered in a particular direction and provides a way to continuously monitor the atomic system detecting the photons that leak through the mirrors \cite{Bux2013,Kessler2014,Landig2015}. 
We model the evolution of the atomic system using the Bose-Hubbard Hamiltonian
\begin{eqnarray}\label{eq:hubbard}
\h{H}_A= -\hbar J \sum_{\langle i,j\rangle}b_j^\dagger b_i +  \frac{\hbar U}{2} \sum_i \hat{n}_i \left(\hat{n}_i-1\right),
\end{eqnarray}
where $b_i$ is the bosonic annihilation operator for the site $i$, $J$ is the tunneling amplitude between neighbor sites and $U$ is the local interaction energy.  The coupling between light and matter is analogous to classical optics where light scattering depends on the overlap between the light mode functions  $u_l(\b{r})$ and the atomic density $\hat{n}(\b{r})=\hat{\Psi}^\dagger(\b{r}) \hat{\Psi}(\b{r})$. Here, we express the matter field operator $\hat{\Psi}(\b{r})$ in terms of the lattice Wannier functions $w(\b{r})$, i.~e. $\hat{\Psi}(\b{r})=\sum_i w(\b{r}-\b{r}_i) b_i$, so that the Hamiltonian that couples light and the atoms is
\begin{eqnarray}\label{eq:coupling}
\h{H}_{LA}= \sum_{l,m}\Omega_{lm}a_l^\dagger a_m \hat{F}_{lm}
\end{eqnarray}
where $\Omega_{lm}=g_l g_m/\Delta_a$, $g_l$ are the atom-light coupling constants, $\Delta_a=\omega_p-\omega_a$ is the probe-atom detuning and
\begin{eqnarray}
\h{F}_{lm} &=\h{D}_{lm}+\h{B}_{lm} =\sum_j J^{lm}_{jj} \h{n}_j +\sum_{\langle i,j \rangle} J^{lm}_{ij} \h{b}^\dagger_i \h{b}_j,\\
J^{lm}_{ij} &= \int\!u^*_l(\b{r}_i)w(\b{r}_i)w(\b{r}_j)u_m(\b{r}_j).
\end{eqnarray}
This model can be also generalized to Fermi systems introducing the light polarization as additional degree of freedom which allows to selectively probe different spin states \cite{Meineke2012, Sanner2012} triggering the formation of antiferromagnetic states \cite{Mazzucchi2015,Kaczmarczyk2016}.
In this work, we consider the case where only two far-detuned light modes are present: the classical coherent probe $a_0$ (c-number) and the cavity mode $a_1$ (operator). Furthermore, we focus on the regime  $\kappa\gg\Delta_p$ where $\Delta_p$ is the cavity-probe detuning so that the effective cavity-mediated interaction can be neglected and the light field affects the atomic state only via measurement backaction. From the coupling Hamiltonian (\ref{eq:coupling}), we can compute the Heisenberg equations describing the evolution of the scattered light. Taking the stationary limit of such equations allows us to adiabatically eliminate the light degrees of freedom \cite{Mekhov2009PRA}. Within this assumptions and neglecting the cavity dispersive shift, annihilation operator $a_1$  can be expressed in terms of the atomic variables , i. e.  $a_1=C \h{F}_{10}$ where $C$ is the Rayleigh scattering coefficient $C=i \Omega_{10} a_0 / (i \Delta_p -\kappa)$. Finally, for deep optical lattices the  operator $\h{D}_{10}$ dominates over $\h{B}_{10}$ so that the light operator describing the photons escaping the cavity is  $a_1=C \h{D}_{10}$. Note that this is not always the case and carefully tailoring the light mode functions it is possible to suppres the contribution form $\h{D}_{10}$ so that the scattered light is sensitive to $\h{B}_{10}$  \cite{Kozlowski,Kozlowski2016}.

The coefficients $J_{jj}$ can be easily engineered changing the light mode function. If $J_{jj}$ has the same value in a region of the optical lattice, atoms in this partition scatter light with the same intensity and phase, making them indistinguishable by the measurement. Importantly, this is a consequence of global coupling between the atoms and the light mode: all the lattice sites scatter light coherently into the cavity.  This is in contrast to recent monitoring schemes that rely on local addressing~\cite{Daley2014,Bernier2014,Hofstetter2014,Hartmann2012,RempeScience2008,LesanovskyPRL2012,LesanovskyPRB2014} 
 where each atom scatter light independently  (i. e. there is a jump operator $\h{c}_j$ for each lattice site) and  the measurement process quickly destroys long-range coherence. If both light modes are traveling waves ($u_l(\b{r}) = e^{i\b{k}_l \cdot \b{r}}$) and considering a one-dimensional chain, the coefficient $J_{jj}$ are analogous to a classical diffraction grating, i. e. $J_{jj}=e^{i \delta j}$ where $\delta$ is the projection of the difference between the wave vectors of the light modes on the direction of the optical lattice. Tuning  the angle between probe and scattered light so that $\delta  = 2 \pi s/R$ ($s, R \in \mathbb{Z}^+$), the optical lattice is partitioned in $R$ macroscopically occupied spatial modes where atoms separated by $R$ lattice sites cannot be distinguished by the measurement. In this geometrical configuration, the dynamics of the system is determined by the evolution of the collective variables characterizing the modes and not by the occupation of each lattice site.
 
We focus on the conditional dynamics in a single experimental realization and we model the evolution of the system with the  quantum trajectories technique \cite{Wiseman}. Within this formalism, the photodetections are described by the application of the quantum jump operator $\h{c}=\sqrt{2 \kappa} \h{a}_1$ (which follows a stochastic process) while the dynamics between two photocounts is determined by the non-Hermitian Hamiltonian
\begin{eqnarray}
\h{H}_{\mathrm{eff}}=\h{H}_A - i \hbar \h{c}^\dagger \h{c}/2.
\end{eqnarray}
The second term of this expression defines an additional energy scale $\gamma=|C|^2 \kappa$ for the system which competes with the usual local processes that describe the evolution of the system (tunneling and on-site interaction).

\section{Effective dynamics of the macroscopic spatial modes}
We start by considering the evolution of a quantum gas with $N$ atoms initially in the superfluid state
\begin{eqnarray}\label{eq: superfluid}
\ket{\Phi(N)}=\frac{1}{\sqrt{M^NN!}}\left( \sum_{i=1}^N b_i^\dagger \right)^N \! \! \ket{0}
\end{eqnarray}
where $\ket{0}$ is the vacuum state for the operators $b_i$. We continuously monitor this system using  traveling waves so that the measurement scheme defines $R$ macroscopically occupied spatial modes and the jump operator is  $\h{c} \propto a_1=C \sum_{j=1}^R e^{i 2 \pi j /R} \h{N}_j$ where $\h{N}_j$ is the occupation of the mode $j$. Making use of the multinomial expansion for the sum in equation (\ref{eq: superfluid}) and assuming that each mode has the same number of lattice sites, we can group the creation operators that operates on the same mode so that $\ket{\Phi(N)}$ can be rewritten as 
\begin{eqnarray}
\ket{\Phi(N)}= \sqrt{\frac{N!}{R^N}}  \sum_{\sum_i N_i=N} \sqrt{\frac{1}{N_1! N_2! \dots N_R!}}  \prod_{i=1}^R \ket{\Phi_i(N_i)}.
\end{eqnarray}
where $\ket{\Phi_i(N_i)}$ is a superfluid in the spatial mode $i$ with $N_i$ atoms. In other words, we decompose a superfluid state in a linear combination of ``smaller'' superfluids that are defined in each spatial mode. This choice is particularly convenient because the states   $\prod_{i=1}^R \ket{\Phi_i(N_i)}$ are eigenvectors of the jump operator
\begin{eqnarray}
\h{c} \prod_{i=1}^R \ket{\Phi_i(N_i)}= \sqrt{2 \kappa} C\left(\sum_{j=1}^R e^{i 2 \pi j /R} N_j \right) \prod_{i=1}^R \ket{\Phi_i(N_i)}.
\end{eqnarray}
Therefore, defining $\mathcal{S}_R$  to be the subspace of the Hilbert space that is spanned by the vectors $\left\{ \prod_{i=1}^R \ket{\Phi_i(N_i)} \right\}_{N_i}$, the dynamics due to the quantum jumps is internal to  $\mathcal{S}_R$. This property enable us to formulate an effective description of the atomic evolution, greatly reducing the computational cost of each quantum trajectory and allowing us to formulate an analytic solvable model. 

Carefully engineering the coefficients $J_{jj}$, we can partition the optical lattice in two spatial modes depending on the parity of the lattice sites. This can be achieved using traveling waves as mode functions for the probe and the cavity (i.~e. $u_l(\b{r})=e^{i \b{k}_l \cdot \b{r}}$) where the wave vectors $\b{k}_0$ and $\b{k}_1$ are orthogonal, corresponding to the detection of the photons scattered in the diffraction minimum and the operator $a_1=C(\h{N}_\even- \h{N}_\odd)$  \cite{MekhovPRL2007,MekhovPRL2009,mekhovLP2009,mekhovLP2010,mekhovLP2011}. Alternatively, one can obtain the same spatial mode structure considering standing waves (i.~e. $u_l(\b{r})=\cos(\b{k}_l \cdot \b{r})$) crossed in such a way that $\b{k}_0$ and $\b{k}_1$ have the same projections on the lattice direction and are shifted in such a way that the even sites of the optical lattice are positioned at the nodes, so that the scattered light operator is $a_1=C\h{N}_\odd$ \cite{Mazzucchi2015}. 
In this case, we can decompose the initial state of the system (superfluid) as 
\begin{eqnarray}\label{eq: superfluid2}
\ket{\Phi(N)}= \sum_{j=1}^N \sqrt{\frac{N!}{2^N j!(N-j)!}} \ket{\Phi_\odd(j),\Phi_\even(N-j)}.
\end{eqnarray}
If the interaction between the atoms can be neglected ($U=0$), the dynamics resulting from the competition between the measurement process and the usual nearest-neighbors tunneling preserves the mode structure and it is possible to describe the system in term of collective variables. To clarify this, we rewrite the tunneling term in equation~(\ref{eq:hubbard}) as 
\begin{eqnarray}\label{eq: partition}
  \sum_{\langle i,j\rangle}b_j^\dagger b_i =  \sum_{i \in \odd} \sum_{j:i} \cop{b}_{j}\aop{b}_i+ \sum_{i \in \even} \sum_{j:i} \cop{b}_{j}\aop{b}_i 
\end{eqnarray}
where $j:i$ indicates that the sites $j$ and $i$ are nearest neighbors. The  dynamics generated by this expression and the non-Hermitian term is internal to the space $\mathcal{S}_2$ if the initial state of the system belongs to $\mathcal{S}_2$. In fact,  by applying $\h{H}_\mathrm{eff}$ to the product of two superfluid states one has
\begin{eqnarray}\label{eq:Heff2}
  \h{H}_\mathrm{eff} & \ket{\Phi_{\odd}(l),\Phi_{\even}(m)}=\nonumber \\ 
  &- \hbar J \sqrt{l(m+1)}\ket{\Phi_{\odd}(l-1),\Phi_{\even}(m+1)}  \nonumber \\
  &- \hbar J \sqrt{m(l+1)}\ket{\Phi_{\odd}(l+1),\Phi_{\even}(m-1)} \nonumber \\
  & -i \hbar \frac{\gamma}{2} |\beta_1 l +\beta_2 m|^2 \ket{\Phi_{\odd}(l),\Phi_{\even}(m)},
\end{eqnarray}
where $\beta_1$ and $\beta_2$ depend on the measurement scheme ($\beta_1=-\beta_2=1$ for probing in the diffraction minimum while $\beta_1=1$ and $\beta_2=0$ if only the odd sites are addressed). 
Therefore, the quantum state of the atoms can be expressed as $\ket{\psi}= \sum_{j=0}^N \alpha_j \ket{\Phi_\odd(j),\Phi_\even(N-j)}$ where $\alpha_j \in \mathbb{C}$ at all times. This allows us  to reformulate the conditional dynamics of the system in term of an effective double-well problem where the occupation of the two wells corresponds to the population of the spatial modes. However, this approach does not allow us to compute any spatial correlations for the system considered: only ``collective'' properties can be calculated.

The generalization of (\ref{eq:Heff2}) to the case of $R$ modes is straightforward: from the spatial structure of the jump operator one can compute the amplitude of the tunneling processes between different spatial modes and build an effective $R-$wells problem. Moreover, the coupling between the effective wells can be tuned changing the spatial structure of the measurement operator, allowing us to go beyond simple double-well systems \cite{Milburn1998,Ruostekoski2001}. For example, considering the case of $R=3$ spatial modes generated by a measurement scheme where the light mode functions are traveling waves and the jump operator is $\h{c}\propto \sum_{j=1}^3 e^{i 2 \pi j/3} \h{N}_j$, one has that the modes alternates across the lattice as $RGBRGB\dots$. Therefore, each lattice site belonging to the mode $R$ is connected to one site of mode $G$ and one of mode $B$ (indexes can be cycled) so that tunneling processes are allowed between all the modes. This reduces the system to a three-wells problem where a specific linear combination of the atomic populations $\h{N}_j$ is monitored while the atoms are free to hop between the three different wells.  Importantly, this is not the only possible way to divide the optical lattice in three modes: if probe and scattered light are standing waves such that $\b{k}_{1,0} \cdot  \b{r} = \pi/4$, the resulting $J_{jj}$ coefficients are $[0,1/2,1,1/2,0,\dots]$ and the modes alternates as $RGBGRGBGR\dots$. In this case, tunneling is not allowed between mode $R$ and $B$ and the coupling between the effective three-wells describing the conditional dynamics must take this into account by forbidding atomic transfer between two of the effective wells. 

\begin{figure}[t!]
\centering
\includegraphics[width=0.8\textwidth]{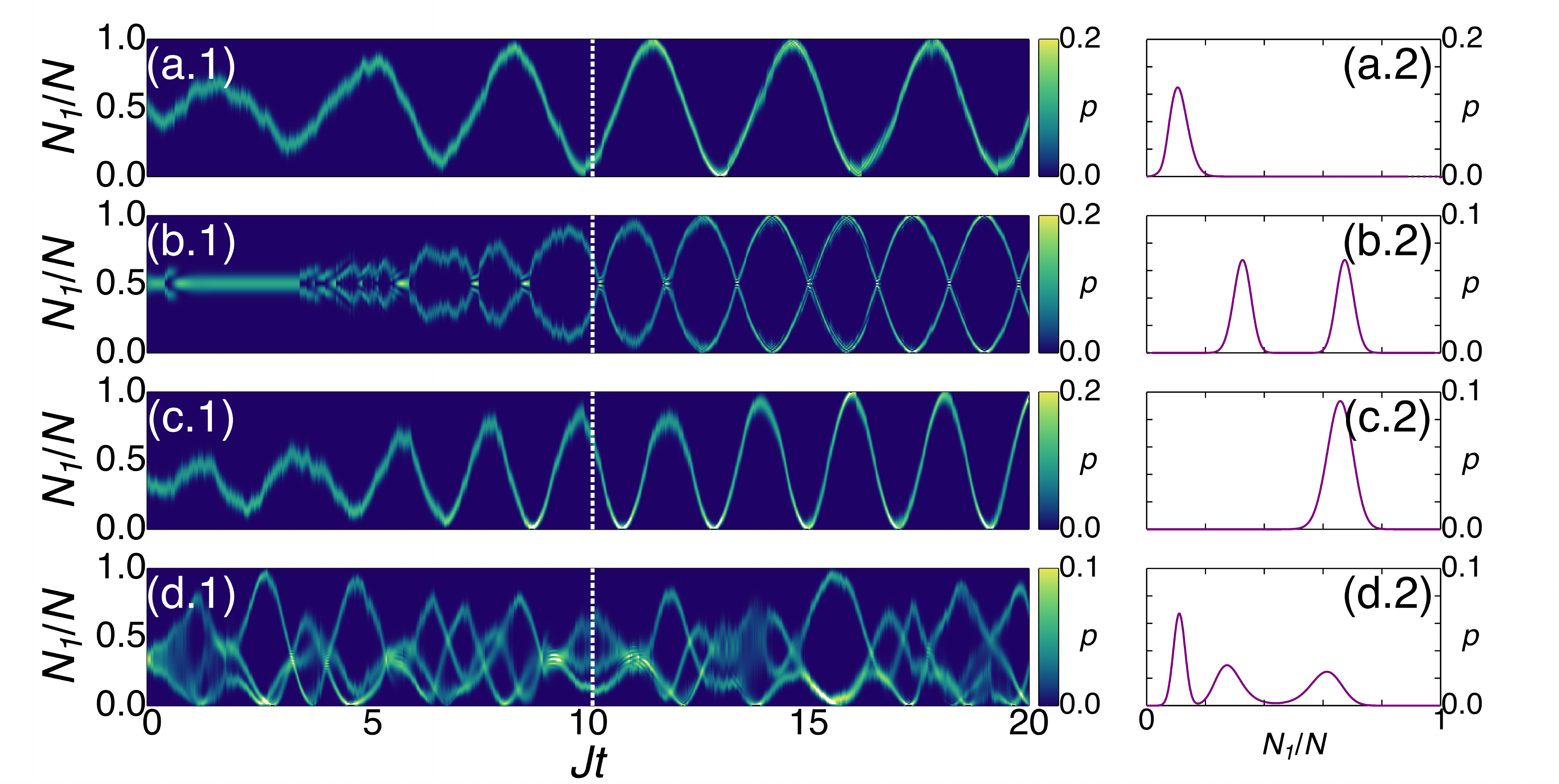}
\caption{ Oscillatory dynamics induce by the measurement process. Probability distribution for the occupation of one of the modes if the optical lattice is partitioned in two (a)-(b) or three spatial modes (c)-(d). Panels (1) illustrate the the full probability distribution for all times while panels (2) focuses on the time indicated by the dashed vertical line.  Depending on the spatial structure of the jump operator, the detection process can lead to multimode macroscopic superpositions. (a) $\gamma/J=0.02, \, N=100, /, J_{jj}=[1,0,1,0,1\dots]$, (b) $\gamma/J=0.02, \, N=100, /, J_{jj}=(-1)^j$, (c) $\gamma/J=0.02, \, N=99, /, J_{jj}=[0,1/2,1,1/2,0\dots]$, (d) $\gamma/J=0.02, \, N=99, /, J_{jj}=e^{i 2 \pi j/3}$ }\label{fig:osc}
\end{figure}

The measurement scheme we consider addresses global atomic observables and therefore preserves long range coherence. Moreover, the spatial structure of the jump operator determines the conditional dynamics: eventual degeneracy of the light intensity $\m{\h{a}_1^\dagger \h{a}_1}$ are imprinted on the state of the atoms. For example, if one detects the photons scattered in the diffraction minimum, so that the measurement addresses the difference in the population of the two spatial modes defined by the odd and even lattice sites, i. e. $\h{a}_1=C (\h{N}_\even -\h{N}_\odd) $, the photon number operator is not sensitive to the sign of such difference. This is because states with opposite $\m{\h{N}_\even -\h{N}_\odd}$ scatter light with different phase but with the same intensity. As a consequence of the measurement backaction, the atomic state becomes a superposition of two macroscopically occupied components: a Schr\"odinger cat state. If the monitoring scheme defines more than two degenerate modes, the conditional evolution leads to an atomic wavefunction that can be expressed as a superposition of multiple macroscopically occupied components. Importantly, this property is a consequence of the spatial structure of the jump operator and does not only depend on the number of modes define by the measurement. For example, considering the case of three modes arranged as $RGBRGB\dots$ ($J_{jj=}e^{i 2 \pi j/3}$), the photon number $\m{\h{a}_1^\dagger \h{a}_1}$ is invariant under the exchange of any two light modes and, as a consequence, the probability distribution of the population of each mode presents three oscillating peaks, indicating that the atomic state conditioned to the measurement is a multimode Schr\"odinger cat state. However, if the spatial modes have a different structure such as $RGBGRGBGR\dots$ ($J_{jj}=[0,1/2,1,1/2,0,\dots]$) this is not the case: only the modes $B$ and $G$ can be exchanged without affecting the intensity of the detected light. Therefore, in this case the probability distribution of the occupation of the mode $R$ has a single peak while the ones for modes $G$ and $B$ are bimodal. This is illustrated in Figure~\ref{fig:osc} where, depending on the measurement scheme, the photodetection induces different dynamics.

The local interaction term in Eq. (\ref{eq:hubbard}) tends to localize the atoms on each lattice site and cannot be included exactly in the approximation we presented. This is because the dynamics described by the interaction operator $ \sum_i \h{n}_i ( \h{n}_i-1)$ is not internal to the subspace $\mathcal{S}_R$. For weak interactions and assuming that the measurement scheme partitions the optical lattice in $R$ spatial modes, we find that the interaction energy $U$ is rescaled by the number of lattice sites belonging to each mode ($M_j$) so that
\begin{eqnarray}
\frac{\hbar U}{2}  \sum_{i=1}^M \h{n}_j ( \h{n}_j-1) \approx \frac{\hbar}{2}  \sum_{j=1}^R \frac{U}{M_j} \h{N}_j ( \h{N}_j-1).
\end{eqnarray}
Note that this expression is approximate and does not allow to describe the strong interacting limit where the atoms form a Mott insulator state. Despite of this limitation, the model we presented allows to describe interacting systems where synthetic interactions mediated by the light field couple different spatial modes \cite{Caballero2016}.

\section{Probing the odd sites of the optical lattice}\label{sec:mf}
If the measurement operator probes the population of the odd sites of the optical lattice, i.~e. $a_1=C\h{N}_\odd$, we can formulate an analytic description of the atomic dynamics~\cite{Julia-Diaz2012}. In between the quantum jumps, the evolution of the system is deterministic and it is determined by the matrix elements of the non-Hermitian Hamiltonian~(\ref{eq:Heff2}):
\begin{eqnarray}\label{eq:mat_el}
\bok{\Phi_\odd(s),\Phi_\even(&N-s)}{\h{H}_\mathrm{eff}}{\psi}=- J \hbar \left( \alpha_{s-1} b_{s-1} + \alpha_{s+1} b_{s}\right)  \nonumber\\ 
&+\frac{2 \hbar U}{M} \left[ s( s-1) + (N-s)(N-s-1) \right] \alpha_s- i \hbar\frac{ \gamma}{2} s^2 \alpha_s
\end{eqnarray}
where $b_s= \sqrt{(s+1)(N-s)}$. Note that this expression does not contain a chemical potential since we will solve the Schr\"odinger equation for a fixed number of particles, as emphasized by the left hand side of equation~(\ref{eq:mat_el}). In the limit $N\gg1$ we can replace the index $s$ with the continuous variable $x=s/N$ which represents the (relative) occupation of the odd sites of the optical lattice. Moreover, we define the  wavefunction $\psi(x=s/N)=\sqrt{N}\alpha_s$ where the $\sqrt{N}$ prefactor ensures that $\psi(x)$  is normalized, i.~e.
\begin{eqnarray}
\sum_{s=0}^N |\alpha_s|^2=\int_{0}^{1} \left| \psi(x)\right|^2 \d  x= 1.
\end{eqnarray}
Introducing $b(x)=\sqrt{(x+h)(1-x)}$, $h=1/N$, $\Lambda=U N /M$ and $\Gamma = N \gamma/2$, and neglecting constant shifts we can rewrite (\ref{eq:mat_el}) as 
\begin{eqnarray}
\bok{x}{\h{H}_\mathrm{eff}}{\psi}=&- \sqrt{N} J \hbar \left[ b(x-h) \psi(x-h)+\psi(x+h)b(x) \right] \nonumber \\
& +2 \sqrt{N}  \hbar \Lambda  x (x-1) \psi(x) - i \sqrt{N} \hbar \Gamma x^2 \psi(x).
\end{eqnarray}
Expanding this expression up to second order in $h$ and defining the normalized atom imbalance $z=(N_\odd-N_\even)/N=2x-1$, we obtain an effective non-Hermitian Hamiltonian that describes the dynamics of the two macroscopically occupied spatial modes. Specifically, we describe the evolution of the system between two quantum jumps with the effective Schr\"odinger equation 
\begin{eqnarray}\label{Schroedinger}
i h \desude{t} \psi(z,t)=H(z) \psi(z,t).
\end{eqnarray}
where the Hamiltonian $H(z)$ is 
\begin{eqnarray}\label{hamiltonian}
H(z)\psi(z)=& -2 J h^2 \desude{z}\left(\sqrt{1-z^2} \desude{z} \psi(z) \right) + \frac{1}{2} \Lambda z^2\nonumber \\
& + V(z) \psi(z)   - i \Gamma \frac{(z+1)^2}{4} \psi(z)
\end{eqnarray}
and the effective potential $V(z)$ is given by
\begin{eqnarray}
V(z)=-J \sqrt{1-z^2} \psi(z) \left[ 1+ \frac{h}{1-z^2}-\frac{h^2(1+z^2)}{(1-z^2)^2}\right].
\end{eqnarray}
The dynamics of the spatial modes is therefore equivalent to the motion of a particle with effective mass $\sqrt{1-z^2}$ in the real potential $V(z)$ and imaginary potential  $- i \Gamma (z+1)^2 /4$. Using the same approximations, we find that the initial state (\ref{eq: superfluid2}) in the limit $N\gg1$ reduces to the Gaussian function
\begin{eqnarray}\label{eq:initial}
\psi(z,0)=\left( \frac{1}{\pi b_0^2} \right)^{1/4} e^{-z^2/(2 b_0^2)}.
\end{eqnarray}
describing a perfectly balanced population between the two spatial modes (i. e. $\m{\h{N}_\odd-\h{N}_\even}=0$) with variance $b_0^2=2h$. In order to give an analytic expression of $\psi(z,t)$, we take the limit of small population unbalance so the mass term becomes $\sqrt{1-z^2}\approx 1$ and we expand the potential $V(z)$ up to second order in $z$:
\begin{eqnarray}
V(z)\approx-1-h + \frac{1}{8}\omega^2 z^2, \qquad \omega=2 \sqrt{1+\Lambda-h}. 
\end{eqnarray}
Therefore, the dynamics of the atomic system is mapped to the evolution of a Gaussian wave packet in an harmonic potential and subjected to dissipation via the non-Hermitian term due to the measurement. Within these assumptions, the wavefunction of the system remains Gaussian at all times and it can be expressed as 
\begin{eqnarray}\label{eq:ansatz}
\psi(z,t)=\left( \frac{1}{\pi b^2(t)} \right)^{1/4} \exp \left[i a(t)+\frac{i z c(t)+ i z^2 \phi(t)-(z-z_0(t))^2}{2 b^2(t)} \right].
\end{eqnarray}
The functions $b^2(t)$, $z_0(t)$, $c(t)$, $\phi(t)$ and $a(t)$ describe the collective dynamics of the system. Specifically, $b^2(t)$ is proportional to the width of the atomic distribution, $z_0(t)$ is the mean value of the unbalance (i. e. $\m{\h{N}_\odd-\h{N}_\even}$) while $c(t)$ and $\phi(t)$ are phase differences between superfluid states with different populations. Moreover, $\operatorname{Re}\left[a(t) \right]$ describes the global phase of the wavefunction and $\operatorname{Im}\left[a(t) \right]$ its norm. Importantly, all these functions are real with the exception of $a(t)$ which is complex. Finally, from the Schr\"odinger equation~(\ref{Schroedinger}) we obtain the differential equations that dictate the evolution of  $b^2(t)$, $z_0(t)$, $c(t)$, $\phi(t)$ and $a(t)$. Specifically, one can prove that
\begin{eqnarray}
	&\dot{(b^2)}=8 h J \phi -  \frac{\Gamma}{2 h} b^4 \label{eq:b1} \\
	&\dot{\phi}=-\frac{J \omega^2}{4h} b^2 - \frac{\Gamma}{ 2 h} b^2 \phi +  \frac{4 h J}{b^2}(1 + \phi^2)\label{eq:phi1} \\
	&\dot{z_0}=- \frac{\Gamma}{2h} b^2 (1 + z_0) + \frac{2 h J}{b^2} (2 z_0 \phi + c)\label{eq:z01} \\
	&\dot{c}=- \frac{\Gamma}{2h} b^2 c +  \frac{4hJ}{b^2}(\phi c - 2 z_0)\label{eq:c1}
\end{eqnarray}
Because of the dissipation, the norm of the wavefunction is not conserved and it is decreasing according to $\exp \left(-2 \operatorname{Im}\left[a(t) \right] \right)$ where 
\begin{eqnarray}
    &\operatorname{Im}\left(\dot{a}\right)=\frac{\Gamma}{4 h} \left[ \left( 1+z_0 \right)^2 + \frac{b^2}{2} \right],
\end{eqnarray}
which determines when a photon escapes the cavity and a quantum jump occurs.

Equations (\ref{eq:b1}-\ref{eq:c1}) can be solved analytically introducing the auxiliary variables  $p=(1- i \phi)/b^2$ and $q=(z_0+i c/2 )/b^2$. Substituting in (\ref{eq:b1}-\ref{eq:c1}), we find that the four equations describing the dynamics of the atomic state reduce to two differential equations:
\begin{eqnarray}
 -2 J h^2 p^2 + \left( \frac{J  \omega^2}{8} - \frac{i \Gamma}{4}\right) + \frac{i h}{2} \frac{\d  p }{\d  t} = 0 \label{eq:p}\\
 4 J h^2 p q - \frac{i \Gamma}{2} - i h\frac{\d  q}{\d  t} =0. \label{eq:q}
\end{eqnarray}
Defining, $\zeta^2=1-i 2 \Gamma / (J\omega^2)$ and making use of standard integrals, one can prove that the solution of the first equation is 
\begin{eqnarray}
p(t)=\frac{\zeta \omega }{4 h} \frac{\left( \zeta \omega + 4 h p(0) \right) e^{i 2 \zeta \omega t} -  \left( \zeta \omega - 4 h p(0) \right)}{\left( \zeta \omega + 4 h p(0) \right) e^{i 2 \zeta \omega t} + \left( \zeta \omega - 4 h p(0) \right)}. \label{eq:p_sol}
\end{eqnarray}
Furthermore, the equation that determines the evolution of $q(t)$ can be solved noting that (\ref{eq:q}) can be rewritten as 
\begin{eqnarray}
\frac{\d }{\d  t} \left( I q \right) = - \frac{\Gamma}{2 h} I.
\end{eqnarray}
where the integrating factor $I$ is given by
\begin{eqnarray}
I&=\exp \left[ i 4 h \int p(t) \d  t \right] \\
&=\left( \zeta \omega + 4 h p(0) \right) e^{i  \zeta \omega t} + \left( \zeta \omega - 4 h p(0) \right) e^{-i  \zeta \omega t}
\end{eqnarray}
so that $q(t)$ is given by
\begin{eqnarray}
q(t)= \frac{1}{2 h \zeta \omega} \frac{A}{ \left( \zeta \omega + 4 h p(0) \right) e ^{i \zeta \omega t} +  \left( \zeta \omega - 4 h p(0) \right) e ^{-i \zeta \omega t}}\label{eq:q_sol} \\
A=  i \Gamma \left[ \left( \zeta \omega + 4 h p(0) \right) e ^{i \zeta \omega t} -  \left( \zeta \omega - 4 h p(0) \right) e ^{-i \zeta \omega t}\right] \nonumber \\
\qquad \qquad +4 h \zeta^2 \omega^2 q(0) -  i 8 h \Gamma p(0) 
\end{eqnarray}
Finally, from these solutions we can extract the physical observables of equations (\ref{eq:b1}-\ref{eq:c1}) as $ b^2(t)=1/\operatorname{Re}\left[ p(t) \right]$,  $ \phi(t)=-\operatorname{Im}\left[ p(t) \right] /\operatorname{Re}\left[ p(t) \right]$, $z_0(t)=\operatorname{Re}\left[ q(t) \right] /\operatorname{Re}\left[ p(t) \right]$ and 
$c(t)=2 \operatorname{Im}\left[ q(t) \right] /\operatorname{Re}\left[ p(t) \right] $.

Instead of focusing on the full solution, here we give a qualitative description of the dynamics generated by these equations. Specifically, we compute the eigenvalues of the Jacobian matrix of the system (\ref{eq:b1}-\ref{eq:c1}) in its stationary points.  Studying the stationary point of the dynamical equations  (\ref{eq:b1}-\ref{eq:c1}) or   (\ref{eq:p}-\ref{eq:q}), we find that there is only one physical critical point. Defining the parameter 
\begin{eqnarray}
\alpha=\sqrt{- \frac{1}{2} +\frac{1}{2} \sqrt{\frac{4 \Gamma ^2}{J^2 \omega ^4}+1}}
\end{eqnarray}
one can prove that 
\begin{eqnarray}
&b^2(\infty)= \frac{4 h J \omega \alpha}{\Gamma} \label{eq:critb}\\
&\phi(\infty)=\frac{J \omega^2 \alpha^2}{\Gamma}\label{eq:critphi}\\
&z_0(\infty)= -1 + \frac{1}{2 \alpha^2+1}\label{eq:critz0}\\
&c(\infty)= \frac{4 \Gamma}{J\omega^2 (2 \alpha^2+1)}.\label{eq:critc}
\end{eqnarray}
Note that these expressions can be obtained also by taking the limit $t\rightarrow \infty$ of the exact solutions (\ref{eq:p_sol}) and (\ref{eq:q_sol}). The eigenvalues of the Jacobian matrix computed in the critical point are are
\begin{eqnarray}
\lambda_{1,2}=\pm \frac{i \Gamma}{\omega \alpha}-J  \omega \alpha   \qquad \mbox{and} \qquad \lambda_{3,4}=\pm \frac{2 i \Gamma}{\omega \alpha}-2 J \omega \alpha.
\end{eqnarray}
Since all of them have a non-positive real part, the point $(b^2,\phi,z_0,c)$ defined by (\ref{eq:critb}-\ref{eq:critc}) is stable. Therefore, the evolution of the system in the long time limit will tend to damped oscillations around the stationary point with frequency $\Omega=2\Gamma/(\omega \alpha)$ and decay time $\Delta t_d=1/(2 J \omega \alpha)$.  The ratio between the measurement strength  $\Gamma$ and the tunneling amplitude $J$ determines if the oscillatory behavior is under- or over-damped. The predictions of this analytic model agree quantitatively with the dynamics described by (\ref{eq:mat_el}) only if the population imbalance between the two spatial modes is small. Despite of this, we find that this simple formulation captures the qualitative behavior of $b^2$ and $z_0$ and help explaining the emergence of the collective oscillations between odd and even sites.

\begin{figure}[t!]
  \centering
  \includegraphics[width=.8\linewidth]{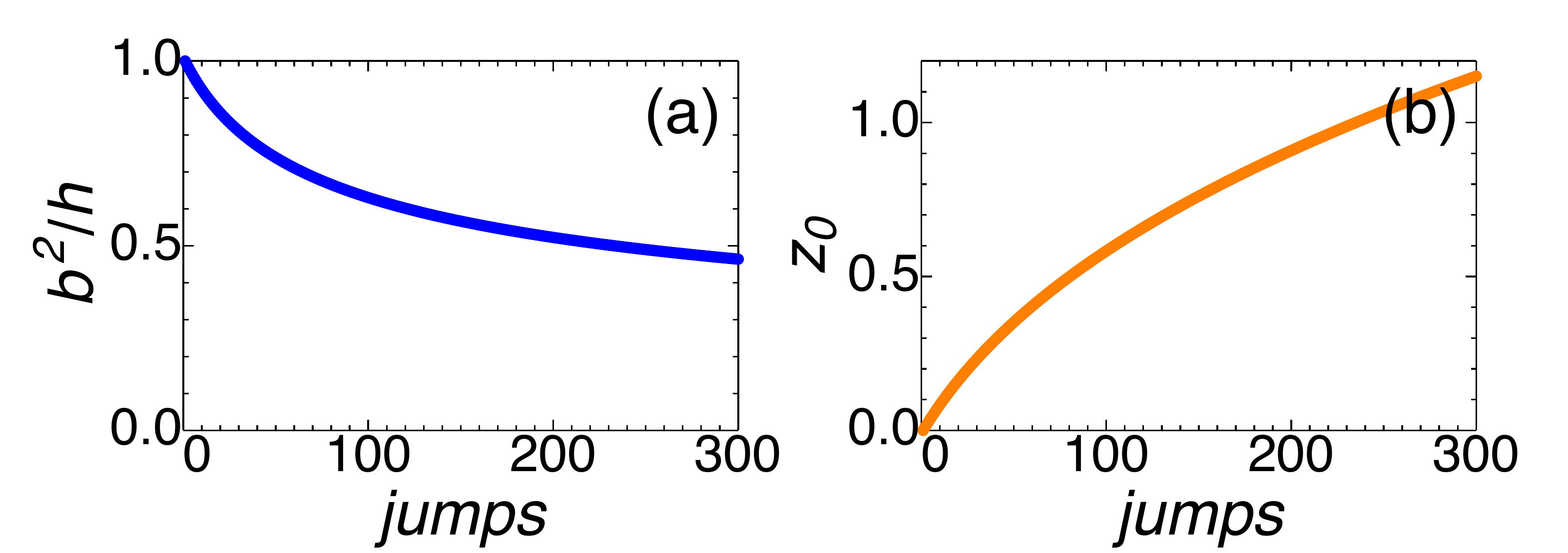}
  \caption{Effect of the quantum jumps on the atomic observables neglecting the effective non-Hermitian dynamics. The uncertainty associated to the number of atoms in each spatial mode decreases (a) while the systems prefers configurations with larger imbalance between odd and even sites (b).}
  \label{fig:jumps}
\end{figure}
The quantum jumps substantially contribute the evolution of the atomic state and drastically alter the dynamics described by (\ref{eq:b1}-\ref{eq:c1}). Their effect can be included in the model by expanding the jump operator $a_1=C(z+1)/2$ around the peak of the Gaussian wavefunction (\ref{eq:ansatz}) as
\begin{eqnarray}
\frac{(1+z)}{2} \approx  \frac{1}{2} \exp\left[ \ln(1+z_0) + \frac{z-z_0}{(1+z_0)} - \frac{(z-z_0)^2}{2(1+z_0)^2}\right].
\end{eqnarray}
Using this expression, we compute the effect of the jumps on the functions $b^2(t)$, $z_0(t)$, $c(t)$ and $\phi(t)$ and we obtain a set of equations that determines the change in initial condition for (\ref{eq:b1}-\ref{eq:c1}) due to the detection of one photon
\begin{eqnarray}
	&b^2 \rightarrow \frac{b^2 (1+z_0)^2}{(1+z_0)^2+b^2} \label{eq:bjump1}\\
	&\phi \rightarrow \frac{\phi(1+z_0)^2}{(1+z_0)^2+b^2}\label{eq:phijump1}\\
	&z_0 \rightarrow z_0 + \frac{b^2(1+z_0)}{(1+z_0)^2+b^2}\label{eq:z0jump1}\\
	&c  \rightarrow  \frac{c(1+z_0)^2}{(1+z_0)^2+b^2}\label{eq:cjump1}.
\end{eqnarray}
Neglecting the non-Hermitian dynamics these equations imply that each quantum jump tends to squeeze the width of the atomic distribution while it increases the atom imbalance between odd and even sites (see~Figure~\ref{fig:jumps}). As a consequence, the measurement process decreases the uncertainty in the population of the spatial modes and the atomic state becomes a product of two superfluid states with well-defined atom number.  In the next paragraphs, we will discuss how the quantum jumps compete with the effective non-unitary dynamics, leading to the creation of states  where the atomic population collectively oscillates between odd and even sites.

\begin{figure}[t!]
  \centering
  \includegraphics[width=.7\linewidth]{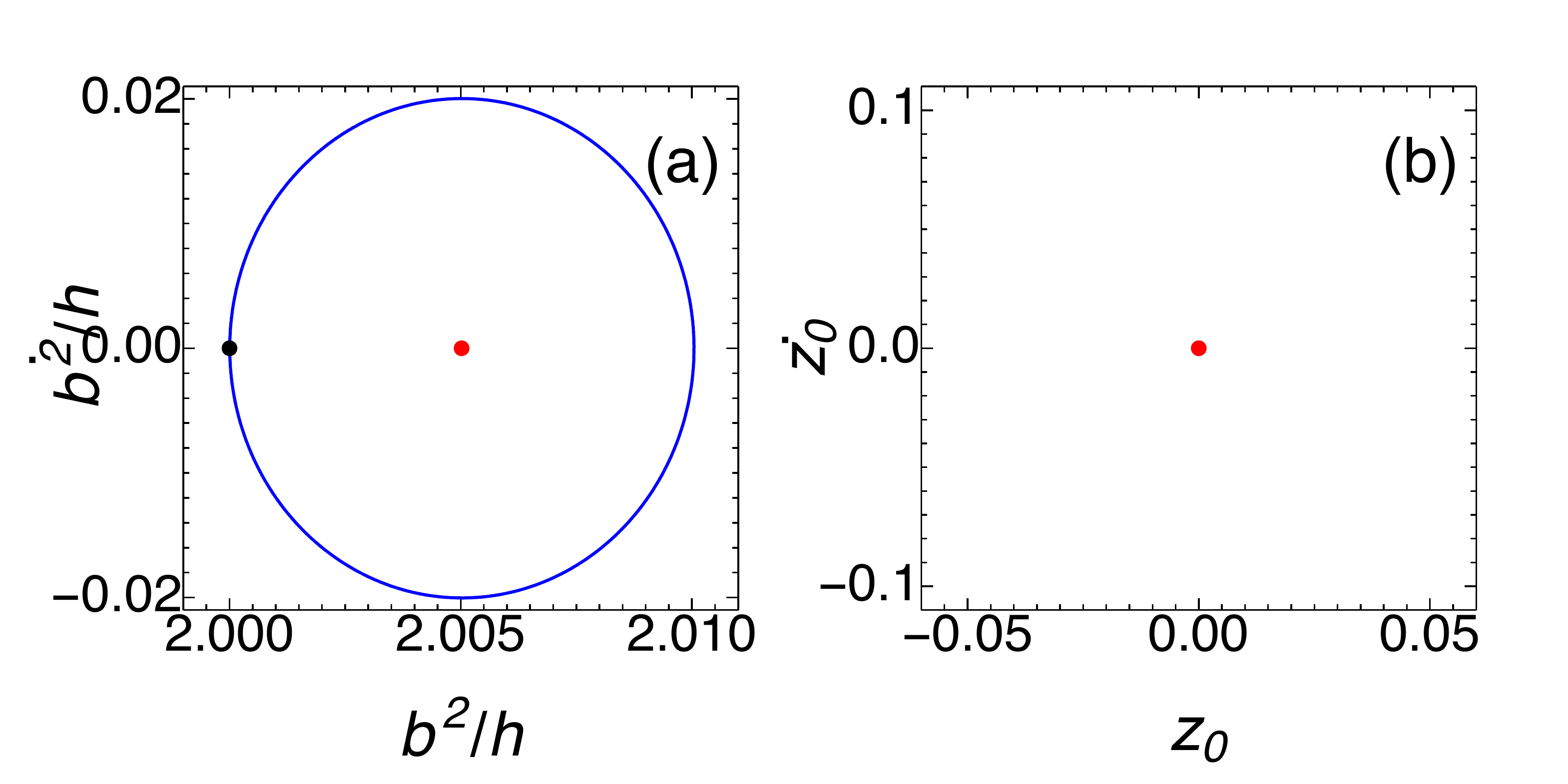}
  \caption{Close orbits around the stable point for the width of the atomic distribution $(b^2,\dot{b^2})$ (a) and the population imbalance $(z_0,\dot{z_0})$ (b) in a single trajectory setting $\Gamma=0$ and without jumps. The black point marks the initial state $(b^2=2 h, \, z_0=0)$ while the red one marks the stationary point. Panel (b) does not show any dynamics since the initial state and the stationary point coincide.}
  \label{fig:hermitianNoJumps}
\end{figure}

\subsection{Case $J=0$}
We first start by considering the case when the atomic tunneling is much slower than the measurement and $J$ can be neglected. Within this assumption, the evolution of the system between two quantum jumps is solely determined by the non-Hermitian dynamics which, together with the quantum jumps, decreases the variance of the population imbalance between odd and even sites. Therefore, the final state of the system is a product of two superfluids with a well-defined number of atoms for each quantum trajectory and the behavior of $b(t)$ is almost deterministic \cite{Onofrio}. However, the imbalance between odd and even sites is not the same for each quantum trajectory since it is determined by the specific sequence of quantum jumps. In fact, the photodetections or the non-Hermitian decay dominates the dynamics of $z_0$ in opposite regimes: the first effect is predominant if the occupation in the odd sites is large while the second one is favored by a large occupation of the even sites.  In each quantum trajectory, these two phenomena balance each other and, in the long time limit, $z_0$ reaches a stationary value which follows the Gaussian probability distribution defined by the initial state (\ref{eq:initial}), favoring states with small population difference between odd and even sites.

\subsection{Case $J\neq0$}
\begin{figure}[t!]
  \centering
  \includegraphics[width=.7\linewidth]{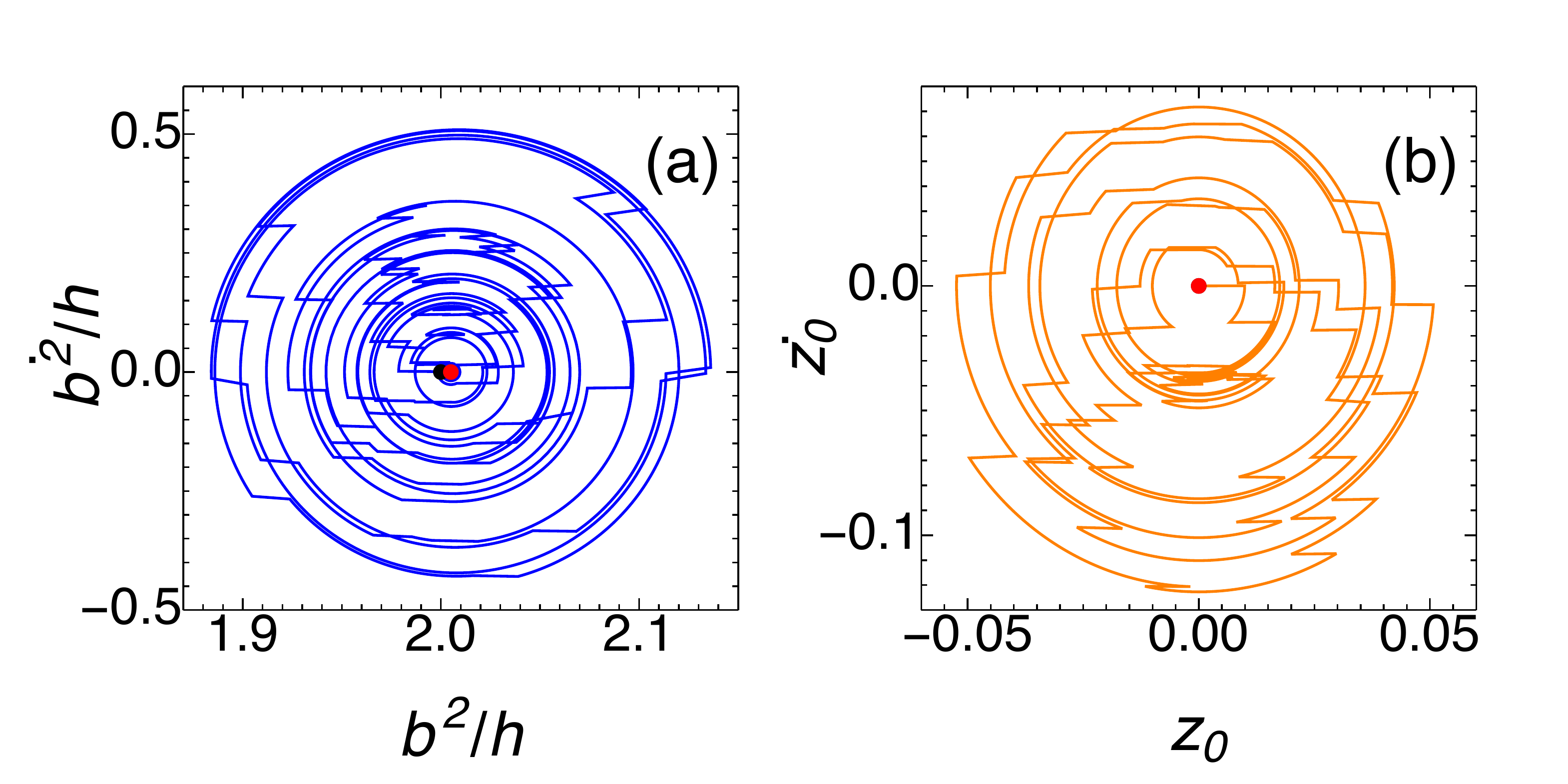}
  \caption{Oscillations of $(b^2,\dot{b^2})$ (a) and $(z_0,\dot{z_0})$ (b) in a single quantum trajectory setting $\Gamma=0$ and applying jumps according to the exact diagonalization solution (\ref{eq:Heff2}). The black point represents the initial state $(b^2=2 h, \, z_0=0)$ while the red one marks the stationary point. The solutions rotate around the stable point with increasing amplitude. Note that the jumps are from right to left for $b^2$ while they are from left to right for $z_0$.}
  \label{fig:hermitianWithJumps}
\end{figure}

\begin{figure}[t!]
  \centering
  \includegraphics[width=.7\linewidth]{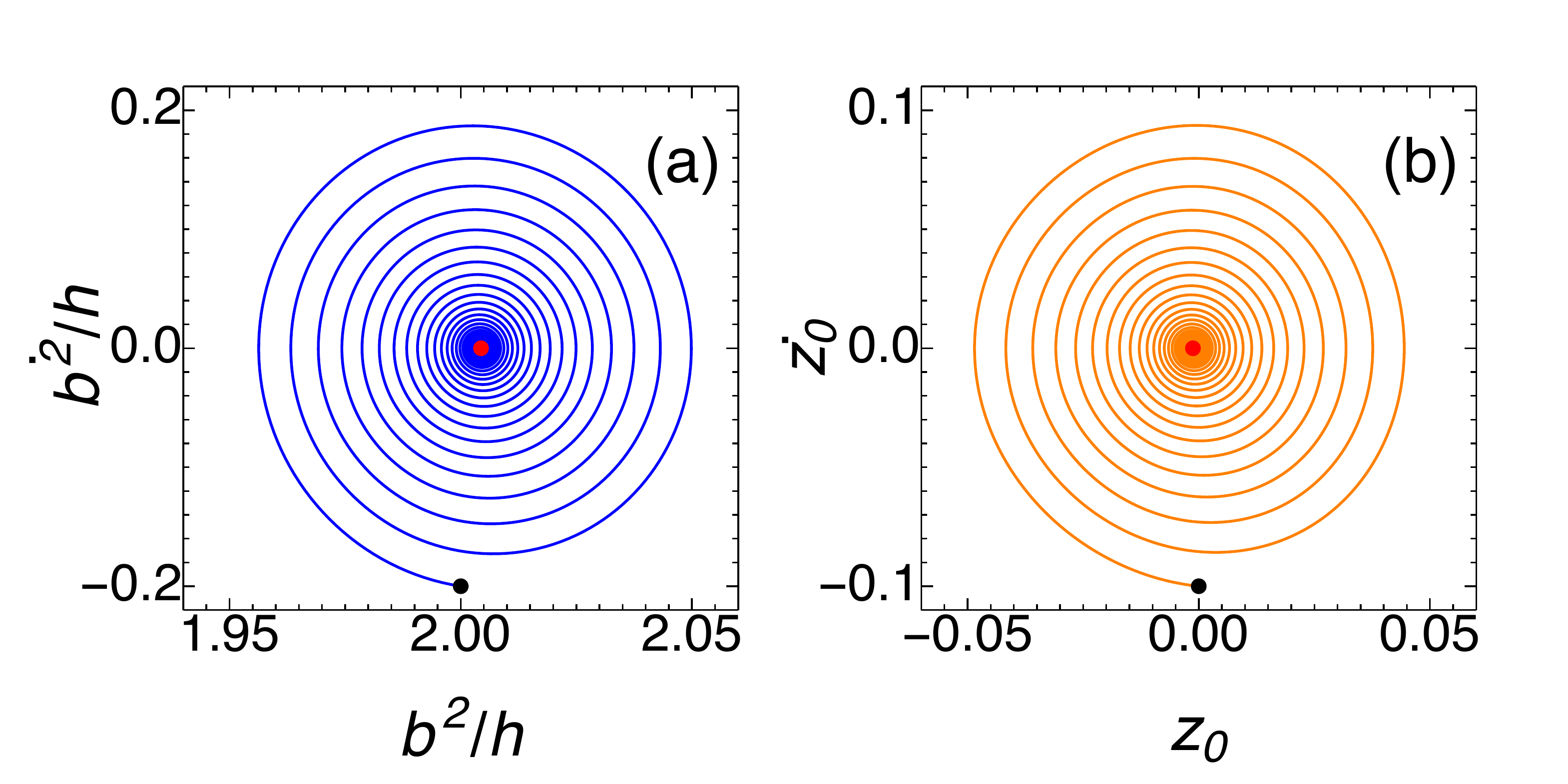}
  \caption{Under-damped oscillations of $(b^2,\dot{b^2})$ (a) and $(z_0,\dot{z_0})$ (b) in a single quantum trajectory in the weak measurement regime ($\Gamma=0.001J$) without quantum jumps. The black point represents the initial state $(b^2=2 h, \, z_0=0)$ while the red one marks the stationary point.}
  \label{fig:weakNoJumps}
\end{figure}

\begin{figure}[t!]
  \centering
  \includegraphics[width=.7\linewidth]{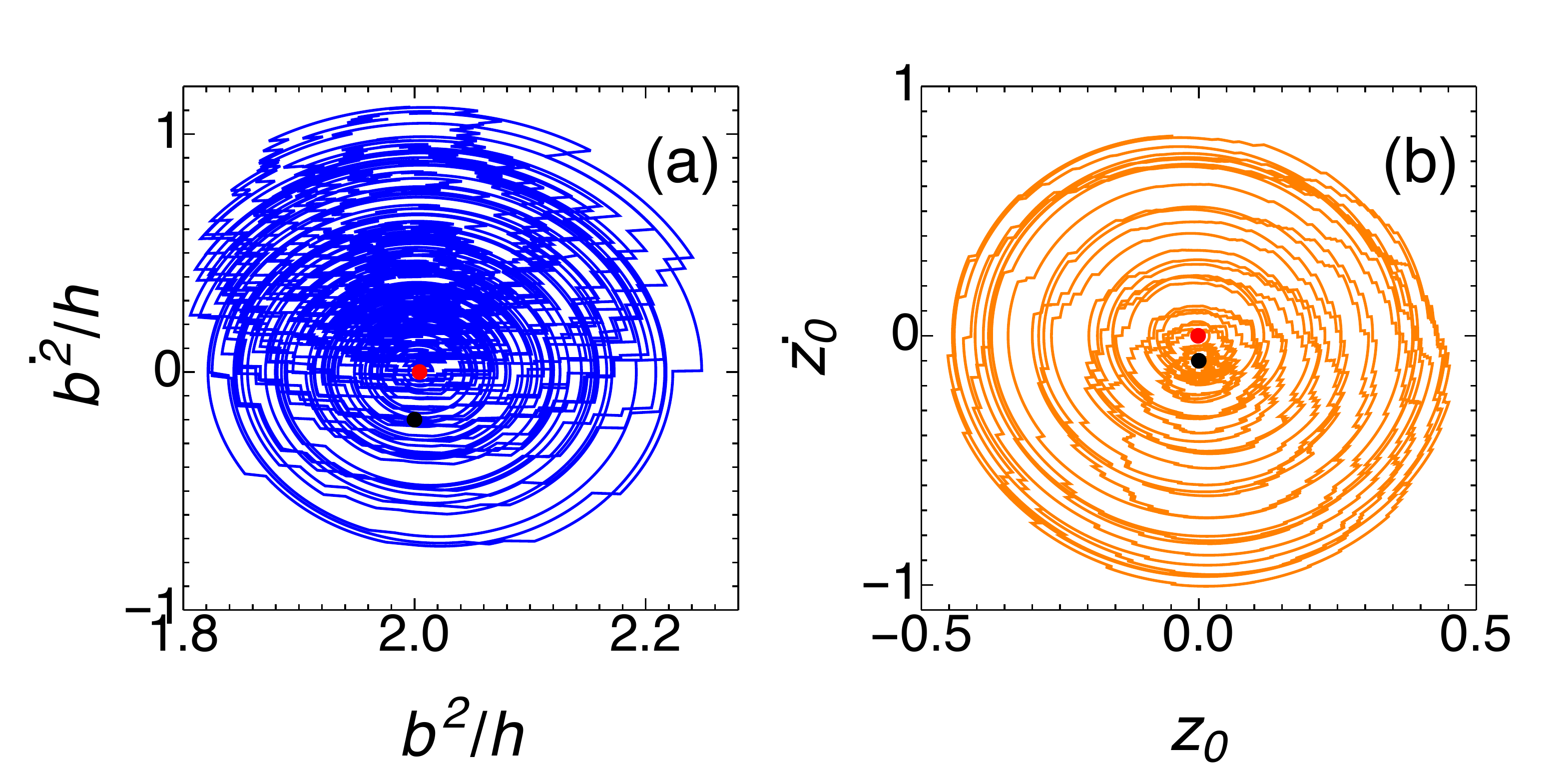}
  \caption{Full conditional dynamics of $(b^2,\dot{b^2})$ (a) and $(z_0,\dot{z_0})$ (b) of a single trajectory in the weak measurement regime ($\Gamma=0.001J$). The black point represents the starting point $(b^2=4 h, \, z_0=0)$ while the red one marks the stationary point.}
  \label{fig:weakWithJumps}
\end{figure}

If the tunneling amplitude $J$ cannot be neglected, the detection process competes with the usual atomic dynamics. 
We first consider the weak measurement limit $\Gamma \ll J$ so that we can describe the evolution of the system between two quantum jumps setting $\Gamma \approx 0$ in (\ref{eq:b1}-\ref{eq:c1}).  From the stability analysis we find that the stable point of the system is $b(\infty)\approx 4h/\omega$ and $z_0(\infty)\approx 0$ while the eigenvalues of the Jacobian matrix are purely imaginary, i.~e. $\lambda_{1,2}=~\pm i J \omega$ and $\lambda_{3,4}=\pm 2 i J \omega$. Therefore, in absence of  quantum jumps, the solutions of (\ref{eq:b1}-\ref{eq:c1}) are oscillating around the stable point without damping (see~Figure~\ref{fig:hermitianNoJumps}). The photodetections perturb this regular oscillations and drive the system quasi-periodically, leading to giant oscillations in the population of the spatial modes. Specifically, the quantum jumps tend to increase the value of $z_0$ according to (\ref{eq:bjump1}-\ref{eq:cjump1}) and consequently, the radius of the oscillations in the $(z_0,\dot{z_0})$ plane is increasing if a jump happens when $z_0>0$ while it is decreasing when $z_0<0$.  Importantly, these two processes do not happen with the same rate because the probability for the emission of a photon in the time interval $\delta t$ depends on the atomic state and it is given by
\begin{eqnarray}\label{prob}
p_{\mathrm{jump}}=\frac{\Gamma}{2 h} \left[ \left( 1+z_0 \right)^2 + \frac{b^2}{2} \right] \delta t.
\end{eqnarray}
Therefore, jumps that increase the radius of the oscillations happen more often and increase the amplitude of the oscillations of $z_0(t)$ (see~Figure~\ref{fig:hermitianWithJumps}).
In order to confirm this prediction, we now turn to the full measurement problem. Taking into account the non-Hermitian dynamics in the differential equations for $b(t)$ and $z(t)$, the radius of the orbits shown in Figure~\ref{fig:hermitianNoJumps} decreases exponentially. Therefore, we can identify three different time scales in the evolution of the system: (i) the oscillation frequency $\Omega=2\Gamma/(\omega \alpha)$, (ii) the damping time $\Delta t_d=1/(2 J \omega \alpha)$ and (iii) the average time interval between two quantum jumps $\Delta t_j=2h/\Gamma$. The ratios between these quantities determine which process is dominating the physics of the system. 
Considering again the weak measurement regime ($\Gamma \ll J$) but taking into account the terms depending on $\Gamma$ in equations (\ref{eq:b1}-\ref{eq:c1}), we find that both $b(t)$ and $z_0(t)$ oscillate around the stationary point with decreasing radius. In this limit, one has $\Omega \Delta t_d \approx J \omega^2/\Gamma \gg 1$, indicating that the system behaves like an under-damped oscillator (Figure~\ref{fig:weakNoJumps}). Importantly, there  are many photocounts during each oscillation ($\Omega \Delta t_j \approx \Gamma h /(J \omega^3) \ll 1$) and the quantum jumps can counteract the damping, driving the atomic system towards states with high population imbalance. In order to prove this, we describe the \emph{average} effect of a quantum jump on the width of the atomic distribution and the relative imbalance as 
\begin{eqnarray}
\delta b^2= \Delta b^2 \, p_{\mathrm{jump}} \qquad \mbox{and} \qquad \delta z_0= \Delta z_0 \, p_{\mathrm{jump}}
\end{eqnarray}
where $\Delta b^2$ and  $\Delta z_0$ are the effect of a single jump on $b^2$ and $z_0$ computed using (\ref{eq:bjump1}) and (\ref{eq:z0jump1}). From these expressions we find that the average photocurrent affects $b^2$ and $z_0$ as
\begin{eqnarray}
&\frac{\delta b^2}{\delta t}= - \frac{\Gamma}{2 h} b^4(t) \left[1- \frac{1}{2}\frac{b^2(t)}{(z_0(t)+1)^2+b^2(t)}\right] \label{eq:mean1}\\
&\frac{\delta z_0}{\delta t}= \frac{\Gamma}{2 h} b^2(t) (z_0(t)+1) \left[1- \frac{1}{2}\frac{b^2(t)}{(z_0(t)+1)^2+b^2(t)}\right] \label{eq:mean2}.
\end{eqnarray}
Note that these equations are consistent with the case $J=0$: the measurement process decreases the width of the atomic distribution and, once $b^2$ reaches its stationary value ($b^2=0$), the unbalance between odd and even sites becomes a constant. 
We compare the exponential damping towards the stable point to the effect of the jumps described by (\ref{eq:mean1}) and (\ref{eq:mean2}). Specifically, solving these equations at first order in $b^2$ we find
\begin{eqnarray}\label{eq:jumps_exp1}
z_{0,\mathrm{jumps}}(t)=-1+(1+z_0(0))\mathrm{e}^\frac{b^2 \Gamma t}{2h}.
\end{eqnarray}
The exponent in this expression should be compared with the one describing the exponential decay of $z_0(t)$. In the weak measurement regime, the evolution between two quantum jumps follows 
\begin{eqnarray}\label{eq:diff_exp1}
z_0(t)=\frac{1}{2} \mathrm{e}^{-\frac{\Gamma}{\omega}t } \left[c(0) \sin (J \omega t)+2 z_0(0) \cos (J \omega t)\right].
\end{eqnarray}
Therefore, the difference between the exponents in (\ref{eq:jumps_exp1}) and (\ref{eq:diff_exp1}) is
\begin{eqnarray}
\Gamma \left(\frac{b^2}{2h}-\frac{1}{\omega}\right),
\end{eqnarray}
which, since $\omega=2 \sqrt{(1-h)}$ and $b^2\sim 2 h$, is positive. This confirms that the jumps increase the amplitude of the oscillations driving the system away from the stable point (Figure~\ref{fig:weakWithJumps}).

In order to estimate the behavior of the imbalance in the large time limit taking into account both the effective dynamics and the quantum jumps, we compute analogous equations to (\ref{eq:mean1}) and (\ref{eq:mean2}) for the phases $\phi$ and $c$, and we incorporate them in the system (\ref{eq:b1}-\ref{eq:c1}). Expanding the resulting expressions at first order in $h$ we find 
\begin{eqnarray}\label{syst}
	&\dot{(b^2)}=8 h J \phi -  \frac{\Gamma}{h} b^4 \\
	&\dot{\phi}=-\frac{J \omega^2}{4h} b^2 - \frac{\Gamma}{ h} b^2 \phi +  \frac{4 h J}{b^2}(1 + \phi^2)\\
	&\dot{z_0}= \frac{2 h J}{b^2} (2 z_0 \phi + c)\\
	&\dot{c}=- \frac{\Gamma}{h} b^2 c +  \frac{4hJ}{b^2}(\phi c - 2 z_0)
\end{eqnarray}
In the large time limit, the width of the atomic distribution becomes constant since the squeezing due to the measurement and the spreading due to the tunneling balance each other so that $\dot{(b^2)}=0$ and $\dot{\phi}=0$. Rearranging the equation for $z_0(t)$ in this limit we find
\begin{eqnarray}
\ddot{z_0}=-\omega^2z_0,
\end{eqnarray}
i. e. the population oscillates between the spatial modes without decaying (Figure~\ref{fig:weakWithJumps}).  

\begin{figure}[t!]
  \centering
  \includegraphics[width=.7\linewidth]{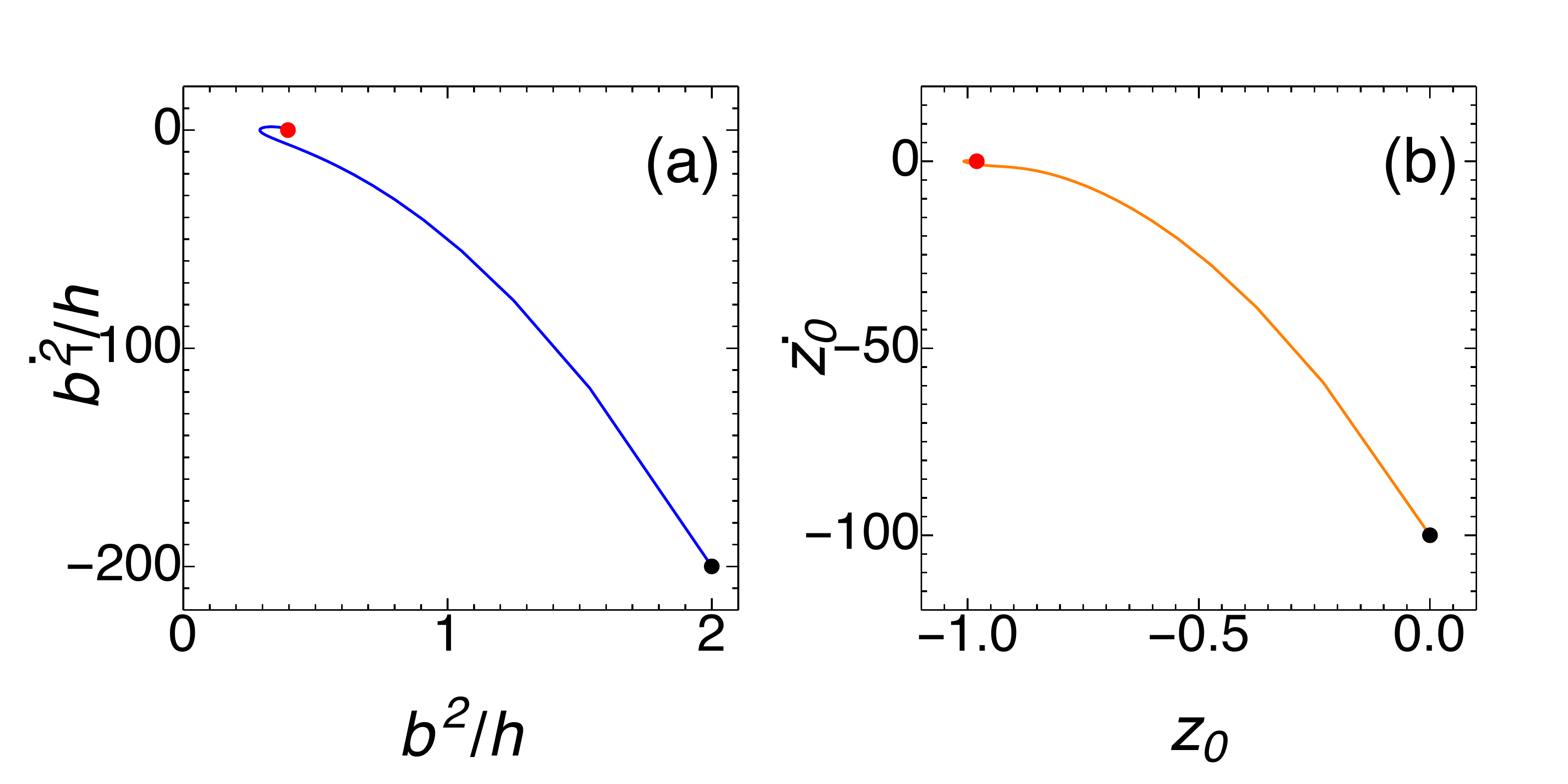}
  \caption{Over-damped oscillations of $(b^2,\dot{b^2})$ (a) and $(z_0,\dot{z_0})$ (b) of a single trajectory in the strong measurement regime ($\Gamma=100J$). The black point represents the starting point $(b^2=4 h, \, z_0=0)$ while the red one marks the stationary point.}
  \label{fig:strongNoJumps}
\end{figure}

\begin{figure}[t!]
  \centering
  \includegraphics[width=.7\linewidth]{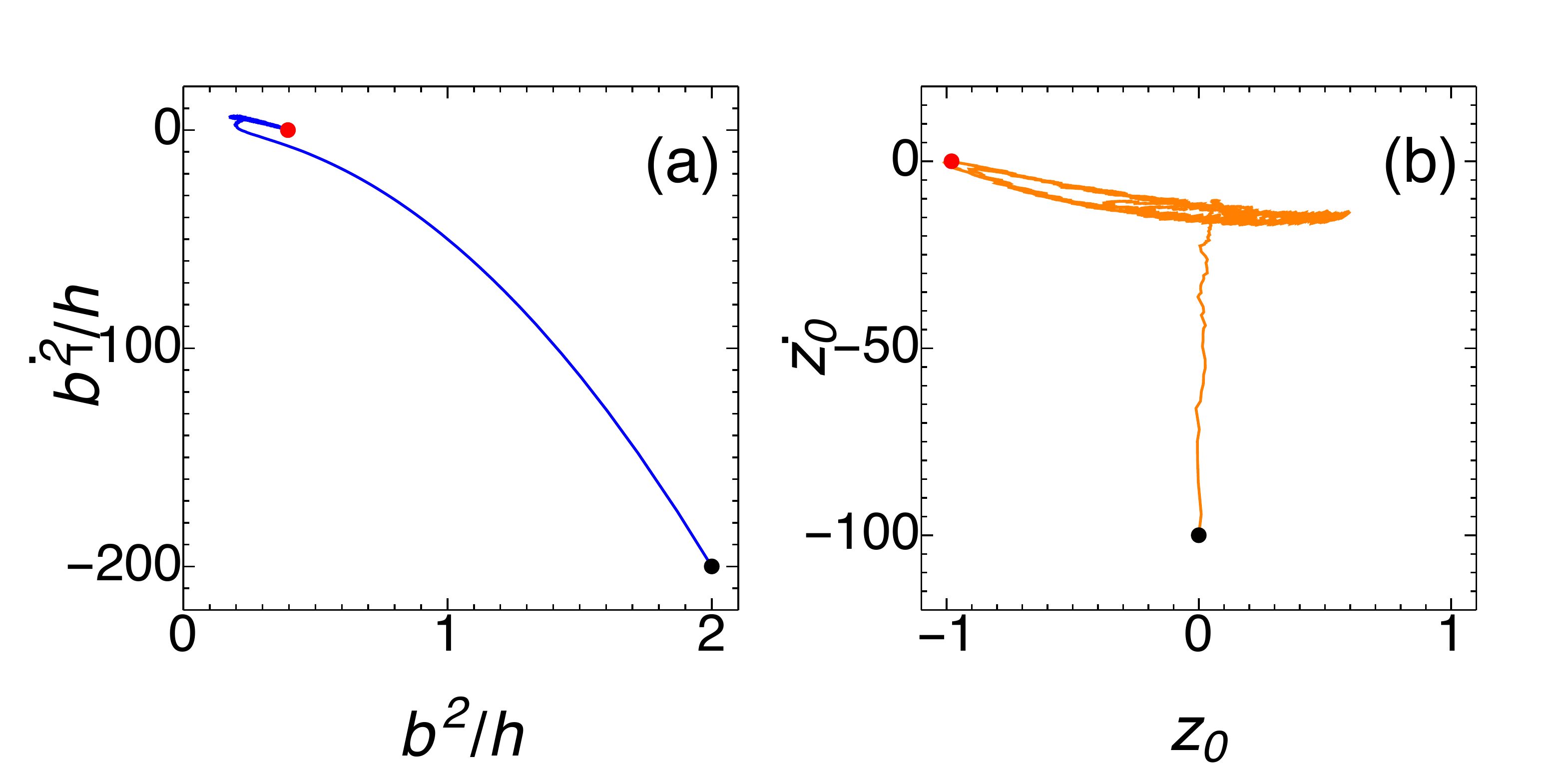}
  \caption{Full conditional dynamics of $(b^2,\dot{b^2})$ (a) and $(z_0,\dot{z_0})$ (b) of a single trajectory in the strong measurement regime ($\Gamma=100J$). The black point represents the starting point $(b^2=4 h, \, z_0=0)$ while the red one marks the stationary point.}
  \label{fig:strongWithJumps}
\end{figure}

If the measurement dominates the dynamics, i.~e. $\Gamma \gg J$,  the non-Hermitian dynamics dominates the evolution between two quantum jumps, In this case, the coordinates of the stationary point in this regime are $b^2(\infty)=4 h \sqrt{J/\Gamma} $ and $z_0 (\infty)=-1+J \omega^2/(2\Gamma)$, i. e. the width of the atomic distribution is extremely squeezed while the odd sites of the lattice tend to be empty.  Importantly, the evolution of the system is not oscillatory since the equations of motions around the stable point resemble an over-damped oscillator as the eigenvalues of the Jacobian matrix are $\lambda_{1,2}=\pm i\sqrt{J \Gamma}- \sqrt{J\Gamma}$ and $ \lambda_{3,4}=\pm 2i\sqrt{J \Gamma}- 2\sqrt{J \Gamma}$. In other words, the period  of an oscillation around the stable point and the damping time are approximately the same ($\Omega \Delta t_{d} \approx 1+ J \omega ^2/(2 \Gamma )$, see Figure~\ref{fig:strongNoJumps}). As we described in the previous paragraphs, the quantum jumps decrease the width of the atomic distribution even further and the full dynamics of $b(t)$ is not qualitatively different from the one determined by the differential equation (\ref{eq:b1}). In contrast, the evolution of the imbalance $z_0(t)$ is heavily affected by the photodetections, as illustrated in Figure~\ref{fig:strongWithJumps}. Specifically, we compare the dynamics due to the quantum jumps (\ref{eq:diff_exp1}) to the one due to the differential equation (\ref{eq:z01}): 
\begin{eqnarray}\label{diff_exp2}
z_0(t)=\frac{1}{2} \mathrm{e}^{-\sqrt{J \Gamma}t } \left[c(0) \sin (\sqrt{J \Gamma}t)+2 z_0(0) \cos (\sqrt{J \Gamma}t)\right].
\end{eqnarray}
Taking the difference between the two exponents we obtain 
\begin{eqnarray}
\Gamma \left(\frac{b^2}{2h}-\frac{1}{\sqrt{J \Gamma}}\right)
\end{eqnarray}
which is always positive, implying that the quantum jumps dominate the dynamics of the system taking $z_0$ away from its stationary point.

\section{Effect of detector efficiency}
\begin{figure}[t!]
\centering
\includegraphics[width=0.9\textwidth]{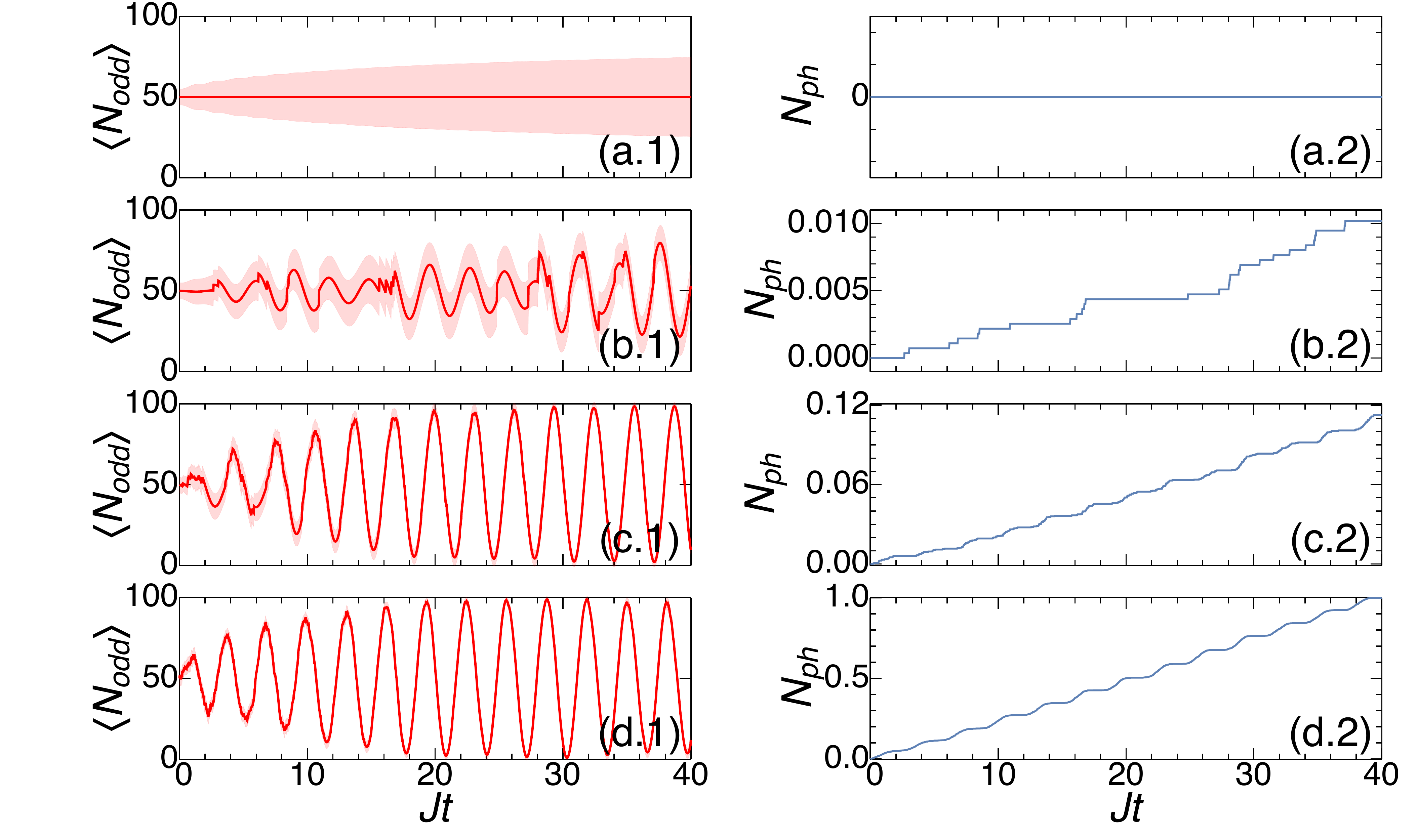}
\caption{ Conditional dynamics measuring $N_{\odd}$ for different detection efficiency obtained solving the SME for different efficiencies ($\eta=0,0.01,0.1,1$ for the panels (a), (b), (c), (d) respectively). Panels (1): atomic population of the odd sites. Panels (2): number of photons detected  $N_{ph}$ (normalized to the case $\eta=1$). The shaded area in the panels (1) represents the fluctuations $\sigma=\sqrt{\mathrm{Tr} \left( \hat{\rho}  \hat{N}_{\odd}^2\right) -\mathrm{Tr} \left( \hat{\rho}  \hat{N}_{\odd}\right)^2}$. Oscillations in the atomic population can be directly observed in the behavior of $N_{ph}(t)$. For zero efficiency (no detector), no oscillations develop, while for finite efficiency the oscillations exist. ($N=100$, $\gamma/J=0.01$)}\label{fig:eff}
\end{figure}

The oscillatory dynamics we presented in the previous sections requires that all the photons leaving the optical cavity are successfully recorded by the detector, i. e. $\eta=1$ where $\eta$ is the detection efficiency. Nevertheless,  the effects we described in this Article can be observed even if $\eta<1$ provided that enough photons are detected for each oscillation period so that it is possible to estimate the photocurrent. Figure~\ref{fig:eff} illustrates this by showing the conditional dynamics of the atomic system for different detection efficiencies when the measurement addresses the population of the odd lattice sites $\h{a}_1=C \h{N}_\odd$.

If the detector is not ideal, it is not possible to describe the conditional dynamics of the system by applying the quantum trajectory technique to the atomic wavefunction. Specifically, if some photons are ``missed'' by the detector, the quantum state of the system becomes mixed and needs to be described by the density matrix $\h{\rho}$ \cite{Wiseman}. Moreover, the evolution of this matrix follows the stochastic master equation (SME)
\begin{eqnarray}\label{sme}
\d \hat{\rho} (t)= \left\{   \d N \mathcal{G} [ \sqrt{\eta}\hat{c} ] - \d t \mathcal{H} [i \hat{H}_0 + \frac{\eta}{2} \hat{c}^\dagger \hat{c} ]  + \d t (1 - \eta)\mathcal{D}[\hat{c}] \right\}\hat{\rho}(t),
\end{eqnarray}
where  $\mathcal{G}, \mathcal{H}$ and  $\mathcal{D}$ are the superoperators 
\begin{eqnarray}
\mathcal{G} [\hat{A} ]\hat{\rho}=\frac{\hat{A}\hat{\rho}\hat{A}^\dagger }{\Tr \left[\hat{A}\hat{\rho}\hat{A}^\dagger \right]}-\hat{\rho}\\
\mathcal{H} [\hat{A} ]\hat{\rho}=\hat{A} \hat{\rho} + \hat{\rho} \hat{A}^\dagger - \Tr\left[ \hat{A} \hat{\rho} + \hat{\rho} \hat{A}^\dagger \right]\\
\mathcal{D} [\hat{A} ]\hat{\rho}=\hat{A} \hat{\rho} \hat{A}^\dagger - \frac{1}{2}\left(  \hat{A}^\dagger  \hat{A} \hat{\rho} + \hat{\rho}  \hat{A}^\dagger  \hat{A}\right).
\end{eqnarray}
and $\d N$ is the stochastic It\^o increment such that $E[\d N]=\eta \Tr[\hat{c} \hat{\rho} \hat{c}^\dagger]\d t$. The physical quantity that is directly accessible in the experiments is $N_{ph}(t)$, i.e.  the number of photons recorded by the detector up to time $t$. Importantly, this function is related to the jump operator and, in the limit where the timescale of the atomic dynamics is much slower than the typical interval between two photocounts, can be expressed as
\begin{eqnarray}\label{der}
\frac{\d N_{ph}}{\d t}=\eta \langle \hat{c}^\dagger \hat{c}\rangle (t),
\end{eqnarray}
where the symbol $\langle \hat{O} \rangle$ represents the expectation value of the operator $\hat{O}$ on a single realization of the SME. Equation (\ref{der}) allows us to estimate the minimum efficiency required for distinguishing the long-range oscillations induced by the measurement backaction. If the population of the odd sites of the optical lattices oscillates in time, the number of photons  recorded by the detector should show a growing ``staircase'' behavior with a characteristic time $2 \pi /J$ (see panel 2 of Figure~\ref{fig:eff} ). We can use this peculiar shape to identify the value of $\m{N_\odd}$: the tread corresponds to the times when $\m{N_\odd} \sim 0$ while the riser coincide with $\m{N_\odd} \sim N$. Therefore, if the detection efficiency allows to clearly resolve each step, the measurement backaction makes the atoms oscillate between odd and even sites retrieving the phenomena that we described in the previous sections. More quantitatively, defining $N_e(t)$ as the number of photons escaping the cavity, the value of $N_{ph}(t)$ follows a Bernoulli process with probability $\eta$ so that  $E[N_{ph}(t)]=\eta N_e(t)$ and $Var[N_{ph}(t)]=\eta (1-\eta) N_e(t)$. Importantly, the detection scheme we consider does not address single site properties since the measurement-induced spatial modes scatter light  collectively. For this reason, the photocurrent $ \eta \langle \hat{c}^\dagger \hat{c}\rangle$ scales as the square of the number of atoms loaded in the optical lattice. This property is crucial and highlights the fact that the oscillatory behavior of the atomic population is not a consequence of the detection of a \emph{single} photon but relies on many collective scattering events. Comparing the variance of $N_{ph}(t)$ to the number of detected photons in a single oscillation period, we estimate that the oscillatory dynamics is present if $\eta\gtrsim J /\gamma  N^2$, making the effects we described robust with respect to detection inefficiency.

\section{Stochastic differential equations and measurement}
In Section~\ref{sec:mf} we investigated the conditional evolution of the atomic system by analyzing the effect of  the non-Hermitian dynamics and the stochastic process described by the quantum jumps separately. However, it is possible to reach the same conclusions by treating these two effects in the same (stochastic) differential equation. Specifically, we model the evolution of the atomic wavefunction in terms of the stochastic Schr\"odinger equation (SSE)~\cite{Wiseman}
\begin{eqnarray}
\d \!\ket{\psi(t)}= &\left[  \d  N(t) \left( \frac{\h{c}}{\sqrt{\m{\h{c}^\dagger \h{c}}(t)}} - \id \right) \right. \nonumber \\
&\qquad+ \left . \d  t \left( \frac{ \m{\h{c}^\dagger \h{c}}(t)}{2} - \frac{\h{c}^\dagger \h{c}}{2} - i  \h{H}\right)\right] \ket{\psi(t)} \label{eq:SSE}
\end{eqnarray}
where $\h{c}$ is the jump operator associated to the measurement, $\h{H}$ is the Hamiltonian generating the coherent dynamics of the system and $\d  N(t)$ is a stochastic increment that obeys the It\^o table
\begin{eqnarray}
&\d  N(t)^2=\d  N(t) \label{eq:Ito1}\\
&\d  N(t)\, \d  t =0.\label{eq:Ito2}
\end{eqnarray}
This SSE describes the atomic dynamics in a single experimental run, i. e. a single quantum trajectory. The stochastic term defines a point process which models the photocounts: if $\d  N(t)=1$ a photon is detected and the quantum jump operator is applied to wavefunction while if $\d  N(t)=0$ the system evolves deterministically. Importantly, the probability of detecting a photon in the (small) time interval $\delta t$ depends on the quantum state of the system and it is given by 
\begin{eqnarray}
p=\bok{\psi(t)}{\h{c}^\dagger \h{c}}{\psi(t)} \delta t.
\end{eqnarray}
In order to give a description of the measurement-induced oscillatory dynamics, we focus on the conditional evolution of the expectation values of few collective variables. From the SSE (\ref{eq:SSE}) and the  It\^o table (\ref{eq:Ito1}-\ref{eq:Ito2}), we find a generalization of the Ehrenfest theorem for the conditional evolution of the observable  $\h{O}$
\begin{eqnarray}\label{eq:SSEobs}
\d \m{\h{O}} (t) =&\left( \frac{\m{\h{c}^\dagger \h{O} \h{c}}}{\m{\h{c}^\dagger \h{c}}} -\m{\h{O}} \right) \d  N(t) \nonumber \\
&+\left(i \m{\left[ \h{H}, \h{O} \right]} - \frac{1}{2}\m{\left\{\h{c}^\dagger\h{c},\h{O} \right\}} + \m{\h{O}} \m{\h{c}^\dagger\h{c}}\right)  \d  t 
\end{eqnarray}
where the expectation values on the right hand side are computed at time $t$ and $[\cdot,\cdot]$ ($\{\cdot,\cdot\}$) is the (anti)commutator. Considering a probe that addresses the population of the odd sites, we can follow the evolution of the system by computing the expectation values of the number of atoms occupying the mode ($\h{N}_\odd$), the atomic current between odd and even sites ($\h{\Delta}$) and their fluctuations $\sigma_{AB}=\m{\h{A}\h{B} + \h{B}\h{A}}/2-\m{\h{A}}\m{\h{B}}$.  From equation (\ref{eq:SSEobs}) we find
\begin{eqnarray}
\d \m{\h{N}_\odd}=&\d  N \left[\frac{\m{\h{N}_\odd^3}}{\m{\h{N}_\odd^2}} - \m{\h{N}_\odd}\right] + \d  t \left[ -J \m{\h{\Delta}} - \gamma \left( \m{\h{N}_\odd^3} -\m{\h{N}_\odd} \m{\h{N}_\odd^2}\right)\right] \label{eq:SSE_dw1}\\
\d \m{\h{\Delta}}=& \d  N \left[\frac{\m{\h{N}_\odd \h{\Delta} \h{N}_\odd}}{\m{\h{N}_\odd^2}} - \m{\h{\Delta}}\right]+ \d  t \left[ -2J \left( N - 2 \m{\h{N}_\odd} \right)  \right. \nonumber \\
&\left.- \frac{ \gamma}{2} \left(\m{\h{N}_\odd ^2 \h{\Delta}} + \m{\h{\Delta} \h{N}_\odd^2} - \m{\h{\Delta}} \m{\h{N}_\odd } \right) \right] \label{eq:SSE_dw2}\\
\d \sigma^2_N=&\d  N \left[\frac{\m{\h{N}_\odd^4}\m{\h{N}_\odd^2}-\m{\h{N}_\odd^3}^2}{\m{\h{N}_\odd^2}} -  \sigma^2_N \right]\nonumber \\
& + \d  t \left[ -J \left( \m{\h{N}_\odd \h{\Delta}} + \m{\h{\Delta} \h{N}_\odd} - 2 \m{\h{\Delta}}\m{\h{N}_\odd}\right) \right. \nonumber \\
& \left. -\frac{ \gamma}{2} \left( 2\m{\h{N}_\odd^4} -4\m{\h{N}_\odd}\m{\h{N}_\odd^3} -2 \m{\h{N}_\odd^2}^2 +4 \m{\h{N}_\odd^2}\m{\h{N}_\odd}^2 \right)\right] \label{eq:SSE_dw3}\\
\d \sigma^2_\Delta=&\d  N \left[\frac{\m{\h{N}_\odd \h{\Delta}^2 \h{N}_\odd}\m{\h{N}_\odd^2}-\m{\h{N}_\odd \h{\Delta} \h{N}_\odd}^2}{\m{\h{N}_\odd^2}} -  \sigma^2_\Delta \right]  \nonumber \\
&+ \d  t \left[ 4J \left( \m{\h{N}_\odd \h{\Delta}} + \m{\h{\Delta} \h{N}_\odd} - 2 \m{\h{\Delta}}\m{\h{N}_\odd}\right) \right.  \nonumber \\
&\left. -\frac{ \gamma}{2} \left( \m{\h{N}_\odd^2 \h{\Delta}^2} + \m{\h{\Delta}^2 \h{N}_\odd^2}-2 \m{\h{\Delta}}(  \m{\h{N}_\odd^2 \h{\Delta}} + \m{\h{\Delta} \h{N}_\odd^2})\right. \right.  \nonumber \\
&\left. \left. - 2 \m{\h{N}_\odd^2}(\m{\h{\Delta}^2}-2\m{\h{\Delta}}^2) \right)\right] \label{eq:SSE_dw4}\\
\d \sigma_{\Delta N}=&\d  N \left[\frac{\m{\h{N}_\odd \h{\Delta} \h{N}_\odd^2}\m{\h{N}_\odd^2}-\m{\h{N}_\odd \h{\Delta} \h{N}_\odd}\m{\h{N}_\odd^3}}{\m{\h{N}_\odd^2}} -  \sigma_{\Delta N} \right] + \nonumber \\ 
&  +\d  t \left[ -J \left( \m{\h{\Delta}^2}-\m{\h{\Delta}}^2 +4 \m{\h{N}_\odd}^2 - 4\m{\h{N}_\odd^2} \right) \right.  \nonumber \\
&\left. -\frac{ \gamma}{2} \left( \m{\h{N}_\odd^2 \h{\Delta} \h{N}_\odd} + \m{\h{\Delta}\h{N}_\odd^3}-2 \m{\h{\Delta}}\m{\h{N}_\odd^3} \right. \right.+ \nonumber \\
& \left.  \left.-\m{\h{N}_\odd}( \m{\h{N}_\odd^2 \h{\Delta}} + \m{\h{\Delta} \h{N}_\odd^2})- 2 \m{\h{N}_\odd^2}(\m{\h{\Delta}\h{N}_\odd}-2\m{\h{\Delta}}\m{\h{N}_\odd}) \right)\right] \label{eq:SSE_dw5}
\end{eqnarray}
where the probability of a jump in a small time interval $\delta t$ is given by
\begin{eqnarray}
p= \delta t \gamma \m{\h{N}_\odd^2}.
\end{eqnarray}
The system (\ref{eq:SSE_dw1}-\ref{eq:SSE_dw4}) is not closed since each equation depends on the expectation values of higher moments of $\h{N}_\odd$ and $\h{\Delta}$. However, we can give an approximate closed formulation of these equations by assuming that  $\m{\h{N}_\odd}$ and $\m{\h{\Delta}}$ are classical Gaussian variables so that all their moments can be expressed as a function of their mean and variance (for example $\m{\h{N}_\odd^3}\approx \m{\h{N}_\odd}^3+3  \m{\h{N}_\odd} \sigma^2_N $. Similarly to Section~\ref{sec:mf}, we focus on the large particle number limit $N\gg 1$ so that it is possible to neglect the variance of $\h{N}_\odd$ with respect to its squared value, i. e. $\sigma^2_N / \m{\h{N}_\odd}^2\sim 1/N \sim 0$. Taking into account these approximations, equations  (\ref{eq:SSE_dw1}-\ref{eq:SSE_dw4}) simplify greatly and can be rewritten as
\begin{eqnarray}
&\d \m{\h{N}_\odd}=  \frac{2 \sigma^2_N}{\m{\h{N}_\odd}}\d  N -  \left( J \m{\h{\Delta}} + 2\gamma \m{\h{N}_\odd}\sigma^2_N\right) \d  t \label{eq:SSE_dw1_2}\\
&\d \m{\h{\Delta}}=  \frac{2 \sigma_{\Delta N}}{\m{\h{N}_\odd}} \d  N -2 \left[ J \left( N-2  \m{\h{N}_\odd}\right) +\gamma \m{\h{N}_\odd} \sigma_{\Delta N} \right] \d  t \label{eq:SSE_dw2_2}\\
&\d \sigma^2_N=-\frac{ 2\sigma^4_N}{\m{\h{N}_\odd}^2}\d  N-2 \left(J  \sigma_{\Delta N} + \gamma \sigma^4_N \right) \d  t \label{eq:SSE_dw3_2}\\
&\d \sigma^2_\Delta =- \frac{ 2\sigma^2_{\Delta N}}{\m{\h{N}_\odd}^2} \d  N  -2 \left(4 J  \sigma_{\Delta N} +\gamma \sigma^2_{\Delta N} \right) \d  t \label{eq:SSE_dw4_2}\\
&\d \sigma_{\Delta N}= \frac{2 \sigma_{\Delta N}\sigma_{N}^2}{\m{\h{N}_\odd}^2} \d  N +  \left[J  (4 \sigma^2_N - \sigma^2_{\Delta}) - 2 \gamma \sigma^2_N\sigma_{\Delta N} \right] \d  t \label{eq:SSE_dw5_2}
\end{eqnarray}
where the jump probability is given by $p= \delta t \gamma  \m{\h{N}_\odd}^2 $.  The deterministic terms in the equations for  $\m{\h{N}_\odd}$ and $\sigma_N^2$, i. e. the ones proportional to the time increment $\d t$, coincide with the differential equations we obtained in Section~\ref{sec:mf} using the mean field approximation. Specifically, we retrieve equations  (\ref{eq:b1}) and (\ref{eq:z01})  by setting $\sigma_N^2=2 N^2 b^2$ and $\m{\h{N}_\odd}=N(1+z_0)$. This confirms that the two different approaches we considered are consistent and lead to the same behavior for the collective variables addressed by the measurement. The main advantage of the system (\ref{eq:SSE_dw1_2}-\ref{eq:SSE_dw5_2}) is that it allows us to describe the quantum jumps and the non-Hermitian dynamics in a single equation. We can use these expressions for gaining insight in the conditional evolution of the atomic system. In order to discuss the behavior of a ``typical'' quantum trajectory, we focus on the equations for the atomic imbalance $\m{\h{N}_\odd}$ and  the current $\m{\h{\Delta}}$ in the limit where the number of photons recorded by the detector can be approximated by a continuous function, i. e. the time interval between two photocounts is much smaller than the  timescale of the atomic dynamics. If this is the case, we can rewrite the It\^o increment as $\d  N=\m{\d  N}-\m{\d  N}+\d  N=\gamma \m{\h{N}_\odd}^2 \d  t + \sqrt{\gamma}\m{\h{N}_\odd} \d  W $ where $ \d  W$ is a Wiener increment representing the fluctuations in the photoncounts around the average value \cite{Wiseman,Ruostekoski2014,Ashida2015b}.  Substituting this expression in (\ref{eq:SSE_dw1_2})-(\ref{eq:SSE_dw1_3}) we find
\begin{eqnarray}
&\d \m{\h{N}_\odd}=  -J \m{\h{\Delta}} \d  t +    2 \sqrt{\gamma} \sigma^2_N \d  W  \label{eq:SSE_dw1_3}\\
&\d \m{\h{\Delta}}=  -2  J \left( N-2  \m{\h{N}_\odd}\right) \d  t  + 2 \sqrt{\gamma} \sigma_{\Delta N} \d  W.  \label{eq:SSE_dw2_3}\\
&\d  \sigma^2_N =  -2  \left( J \sigma_{\Delta N} + 2 \gamma \sigma_N^4 \right) \d  t - \frac{2 \sqrt{\gamma} \sigma_{N}^4}{\m{\h{N}_\odd}} \d  W.  \label{eq:SSE_dw3_3}\\
&\d  \sigma_{\Delta}^2 =  -4  \left( 2J \sigma_{\Delta N} +  \gamma \sigma_{\Delta N}^2 \right) \d  t - \frac{2 \sqrt{\gamma} \sigma_{\Delta N}^2}{\m{\h{N}_\odd}} \d  W.  \label{eq:SSE_dw4_3}\\
&\d  \sigma_{\Delta N} =   J \left( 4\sigma_{N}^2- \sigma_{\Delta}^2  \right) \d  t +\frac{2 \sqrt{\gamma} \sigma_{\Delta N}  \sigma_{N}^2}{\m{\h{N}_\odd}} \d  W.  \label{eq:SSE_dw5_3}
\end{eqnarray}
Neglecting the fluctuations in the photocounts, the equations for $ \m{\h{N}_\odd}$ and  $\m{\h{\Delta}}$ describe the evolution of an harmonic oscillator and confirm the emergence of the oscillatory behavior for the population of the odd sites of the lattice. Note that these oscillations are present even without measurement but here their behavior is fundamentally different: in absence of continuous monitoring the amplitude of the oscillations is proportional to the atom imbalance of the initial state and its probability distribution tends to spread, i. e. the value of $\sigma_N^2$ increases in time. In contrast, here we observe that the uncertainty in the occupation of the spatial modes ($\sigma_N^2$) decreases in time (as suggested by equation ($\ref{eq:SSE_dw3_3}$)) and that full-exchange of atoms between the two spatial modes is possible even starting with a perfectly balanced state. 
An alternative formulation of equations (\ref{eq:SSE_dw1_3})-(\ref{eq:SSE_dw5_3}) can be obtained by rewriting them using the Stratonovich formalism:
\begin{eqnarray}
\frac{\d }{\d  t} \m{\h{N}_\odd}= - J \m{\h{\Delta}} - \gamma \frac{\d }{\d  t} \sigma_N^4 +  2 \sqrt{\gamma} \sigma^2_N \xi (t)\label{eq:strat1}\\
\frac{\d }{\d  t} \m{\h{\Delta}} = -2  J \left( N-2  \m{\h{N}_\odd}\right)  -\gamma \frac{\d }{\d  t} \sigma_{\Delta N}^2+ 2 \sqrt{\gamma} \sigma_{\Delta N} \xi (t)\label{eq:strat2}\\
\frac{\d }{\d  t} \sigma_N^2 =  \frac{-2  \left( J \sigma_{\Delta N} + 2 \gamma \sigma_N^4 \right) + \gamma \sigma_N^4 \frac{\d }{\d  t} \frac{1}{\m{\h{N}_\odd}^2}-\frac{2 \sqrt{\gamma} \sigma_{N}^8}{\m{\h{N}_\odd}} \xi(t)}{1-2 \gamma  \frac{\sigma_{N}^2}{\m{\h{N}_\odd}^2}}\label{eq:strat3}\\
\frac{\d }{\d  t} \sigma_\Delta^2=-4  \left( 2J \sigma_{\Delta N} +  \gamma \sigma_{\Delta N}^2 \right)  + \gamma \frac{\d }{\d  t} \frac{\sigma_{\Delta N}^4}{\m{\h{N}_\odd}^2}- \frac{2 \sqrt{\gamma} \sigma_{\Delta N}^2}{\m{\h{N}_\odd}} \xi(t)\label{eq:strat4}\\
\frac{\d }{\d  t} \sigma_{\Delta N} =   \frac{J \left( 4\sigma_{N}^2- \sigma_{\Delta}^2  \right) -  \gamma \sigma_{\Delta N} \frac{\d }{\d  t} \frac{ \sigma_{N}^4}{\m{\h{N}_\odd}^2}+ \frac{2 \sqrt{\gamma} \sigma_{\Delta N}  \sigma_{N}^2}{\m{\h{N}_\odd}} \xi(t)}{1+\gamma \frac{\sigma_{N}^4}{\m{\h{N}_\odd}^2 }}\label{eq:strat5}
\end{eqnarray}
where $ \xi(t)$ is a Wiener process. Combining Eq. (\ref{eq:strat1}) and (\ref{eq:strat2}), we find that the dynamics of the number of atoms in the odd sites can be described as  a forced harmonic oscillator:
\begin{eqnarray}
\frac{\d ^2}{\d  t^2} \m{\h{N}_\odd}=  2  J^2 \left( N-2  \m{\h{N}_\odd}\right) + F
\end{eqnarray}
where the forcing term is given by
\begin{eqnarray}
F=\gamma \left( J \frac{\d }{\d  t} \sigma_{\Delta N}^2 - \frac{\d ^2}{\d  t^2} \sigma_N^4 \right) + 2 \sqrt{\gamma} \left [ \frac{\d }{\d  t} \left( \sigma_N^2 \xi(t) \right)- \sigma_{\Delta N} \xi(t)\right].
\end{eqnarray}
Therefore, the measurement introduces as a quasi-periodic stochastic force $F$ that drives the system towards larger imbalance, increasing the amplitude of the oscillations of  $ \m{\h{N}_\odd}$.

\section{Extensions for multimode photonic systems}

In this paper we focused on multimode dynamics of ultracold atoms. However, it is reasonable to ask a question, whether the idea of combining the multimode unitary dynamics and quantum backaction of measurement can be extended to other systems.  Recently, significant effort has been made in the development of purely photonic systems with multiple path interferometers, which are one of the setups promising for applications in quantum technologies. A possible realization consists of multiple interconnected fibers, the so called photonic circuits or photonic chips \cite{Spring2013, Holleczek2015}. Indeed, quantum walks \cite{Elster2015} have been already discussed in the contexts of both ultracold atoms in optical lattices and single photons propagating and interfering in a multiple path interferometer. Both systems are the candidates for realizations of quantum simulations and quantum computation protocols.  

The tunneling of atoms in an optical lattice can be analogous to the propagation of photons in the waveguides and their transmission and reflection at beamsplitters (waveguide couplers). Already current technologies allow using single photons and photon pairs as input states for multiple waveguides \cite{Spring2013}.  The boson sampling is considered as a realization of essentially multimode quantum interference of bosons during their unitary evolution (i.e. the propagation through the photonic system).

The detection of photons can be considered in several ways. On the one hand, the non-destructive detection of photons is indeed very difficult to implement. Nevertheless some research is being carried out, which makes it reasonable to expect at least some progress in the future. First, the parametric down conversion produces pairs of entangled photons or beams. The detection of an idler beam represents a QND measurement of the signal beam~\cite{Wiseman}. The use of photon pairs is already consistent with current systems~\cite{Spring2013}. Second, a QND method of photodetection was realized using a cavity QED system~\cite{Reiserer2013}. It indeed remains a challenge to integrate various elements together. On the other hand, as several photons participate in the interference, even the standard detection of a small number of them, while being destructive, can be also considered as a non-fully projective measurement as the rest of photons continue to evolve after some photons are detected. For example, the photon subtraction technique was already shown to produce interesting nonclassical states of light and quantum correlations~\cite{Paternostro2011}. One can draw an analogy with ultracold atoms for the case of two BECs: when the small number of atoms is destructively detected after the matter-wave interference, the remaining atoms develop the phase coherence between two condensates due to the projection of quantum state~\cite{Castin1997}. Thus, while being an experimental challenge, the fully photonic realization of the competition between measurement backaction and unitary dynamics can lead to interesting developments in quantum technologies.

\section{Conclusions}

We have shown that light scattering from ultracold gases in optical lattice can be used for partitioning the system into macroscopically occupied spatial modes with non-trivial overlap which preserve long-range coherence. We formulated an effective model for the dynamics of this mode in a single quantum trajectory, mapping each spatial mode to a single ``well'' and describing its properties in term of collective variables.  The measurement backaction competes with the standard local dynamics and induces oscillatory dynamics on the atomic state. Depending on the spatial profile of the measurement operator, this competition can be exploited for creating multimode macroscopic superposition states which could have applications in metrology and quantum information. Importantly, these states are robust with respect to detection inefficiencies because of the global addressing of our measurement setup. We presented an analytic model that captures the emergence of large-scale collective oscillations with increasing amplitude for the case where only two spatial modes are present. Using the quantum trajectory formalism, we found that  the measurement backaction drives the system away from its stationary point and behaves as a quasi-periodic force acting on the atoms. Finally, we confirmed our finding by formulating an alternative description in terms of stochastic differential equations for the evolution of collective atomic observables. In the limit where the time interval between two photocounts is much smaller than the  timescale of the atomic dynamics and the fluctuations in the photocount can be neglected, the atomic population of one of the modes evolve as an harmonic oscillator driven by a stochastic force.

The measurement scheme described in this work has been recently realized \cite{Hemmerich2015,Esslinger2015} and the phenomena that we predicted can be observed in these setups. Several experiments succeeded in implementing part of our proposal:  light scattered from ultracold atoms in an optical lattice (without a cavity) has been detected~\cite{KetterlePRL2011,Weitenberg2011} and ultracold bosons have been trapped in an optical cavity (without an optical lattice) \cite{EsslingerNat2010,HemmerichScience2012,ZimmermannPRL2014}. 
Moreover, the effect of measurement backaction on atomic system has been observed in the context of 
single-atom \cite{Vengalattore} and multi-particle quantum Zeno
effect \cite{Barontini2015}. Because it relies on off-resonant scattering, the monitoring scheme we propose is not sensitive to the detailed level structure of the system. This makes it applicable to a vast range of quantum objects such as molecules (including biological ones)~\cite{MekhovLP2013}, ions \cite{Blatt2012}, atoms in multiple cavities~\cite{Hartmann2006,Palacios2010,Paraoanu2011,Pirkkalainen2013,White2015,Bretheau2015}, semiconductor~\cite{Trauzettel2007}, multimode cavities \cite{Gopalakrishnan2009,Kollar2015}  or superconducting~\cite{Fink2009} qubits.

\section*{Acknowledgments}
The work was supported by the EPSRC (DTA and EP/I004394/1).

\section*{References}
\bibliography{references}
\bibliographystyle{iopart-num}

\end{document}